\documentclass[prb,twocolumn,showpacs,amsmath,amssymb,superscriptaddress,final]{revtex4}
\usepackage{graphicx} 
\usepackage{dcolumn} 
\usepackage{bm} 
\usepackage{times}
\usepackage[pdftex,colorlinks,bookmarks=false,citecolor=blue,linkcolor=red,urlcolor=blue]{hyperref}
\usepackage{color}
\usepackage{rotating}

\begin{document}

\title{Exact diagonalization study of the antiferromagnetic spin-$\frac{1}{2}$ Heisenberg model\\ on the square lattice in a magnetic field}

\author{Andreas L\"uscher}
\affiliation{Institut Romand de Recherche Num\'erique en Physique des Mat\'eriaux (IRRMA), EPFL, CH-1015 Lausanne, Switzerland}
\author{Andreas M. L\"auchli}
\affiliation{Max-Planck-Institut f\"ur Physik komplexer Systeme, N\"othnitzer Str. 38, D-01187 Dresden, Germany}

\date{\today}

\begin{abstract}
We study the field dependence of the antiferromagnetic spin-$\frac{1}{2}$ Heisenberg model on the square lattice by means of exact diagonalizations. In a first part, we calculate the spin-wave velocity $c$, the spin-stiffness $\rho_s$, and the magnetic susceptibility $\chi_\perp$ and thus determine the microscopic parameters of the low-energy long-wavelength description. In a second part, we present a comprehensive study of dynamical spin correlation functions for magnetic fields ranging from zero up to saturation. We find that at low fields, magnons are well defined in the whole Brillouin zone, but the dispersion is substantially modified by quantum fluctuations compared to the classical spectrum. At higher fields, decay channels open and magnons become unstable with respect to multi-magnon scattering. Our results directly apply to inelastic neutron scattering experiments.
\end{abstract}


\maketitle


\section{Introduction}
The antiferromagnetic spin-$\frac{1}{2}$ Heisenberg model on the square lattice has been investigated in great detail over the past two decades, primarily because the parent compounds of superconducting copper-oxide materials are well described by this model~\cite{manousakis91,barnes91}. On the square lattice, and more generally on any sufficiently connected bipartite lattice, the model is magnetically ordered at zero temperature and the long-wavelength properties such as the magnetic order~\cite{vaknin87}, the low-energy excitations~\cite{coldea01} or the temperature dependence of the magnetic correlations~\cite{birgeneau99,elstner95,cuccoli97} are well described within a semiclassical setting~\cite{anderson52,oguchi60,chakravarty89}. At zero field, the staggered moment $m_s$ is however reduced by around $40\%$ compared to its classical value, indicating the presence of sizable quantum fluctuations. One prominent effect of fluctuations, the renormalization of the spin-wave dispersion along the magnetic zone boundary between momenta $(\pi,0)$ and $(\pi/2,\pi/2)$, was analyzed in Refs.~\onlinecite{kim99,sandvik01,ronnow01,christensen04,christensen07}. Their numerical and experimental findings showed that while the classical theory predicts a flat band along the boundary, the energy and the quasi-particle weight at $(\pi,0)$ are substantially lower than at $(\pi/2,\pi/2)$. 

In the presence of a magnetic field, fluctuations are expected to be even more important because of the additional interaction between spin waves and the field. Since the majority of studies of the Heisenberg model is devoted to the zero or small field situation, these effects are not yet well understood. The negligence of the strong field regime is not surprising - until recently, the prospects of experimentally probing the intermediate or even high field physics were dire, given the huge exchange coupling of layered cuprate materials. This situation has changed with the synthesis of CuBr$_4$ and CuCl$_4$ compounds~\cite{woodward02,lancaster07,coomer07} that have experimentally accessible saturation fields of less than 25 Tesla. In contrast, the high-T$_c$ parent compounds have enormous saturation fields of several thousand Teslas. At low fields, spin-wave theory predicts only a weak renormalization of the magnon dispersion due to hybridization between single- and two-magnon states~\cite{osano82}. Zhitomirsky and Chernyshev~\cite{zhitomirsky99} showed that the effects of this hybridization are more drastic at higher fields, leading to the instability of spin waves with respect to spontaneous decay.
\begin{figure}
\centerline{\includegraphics[width=0.95\linewidth,clip]{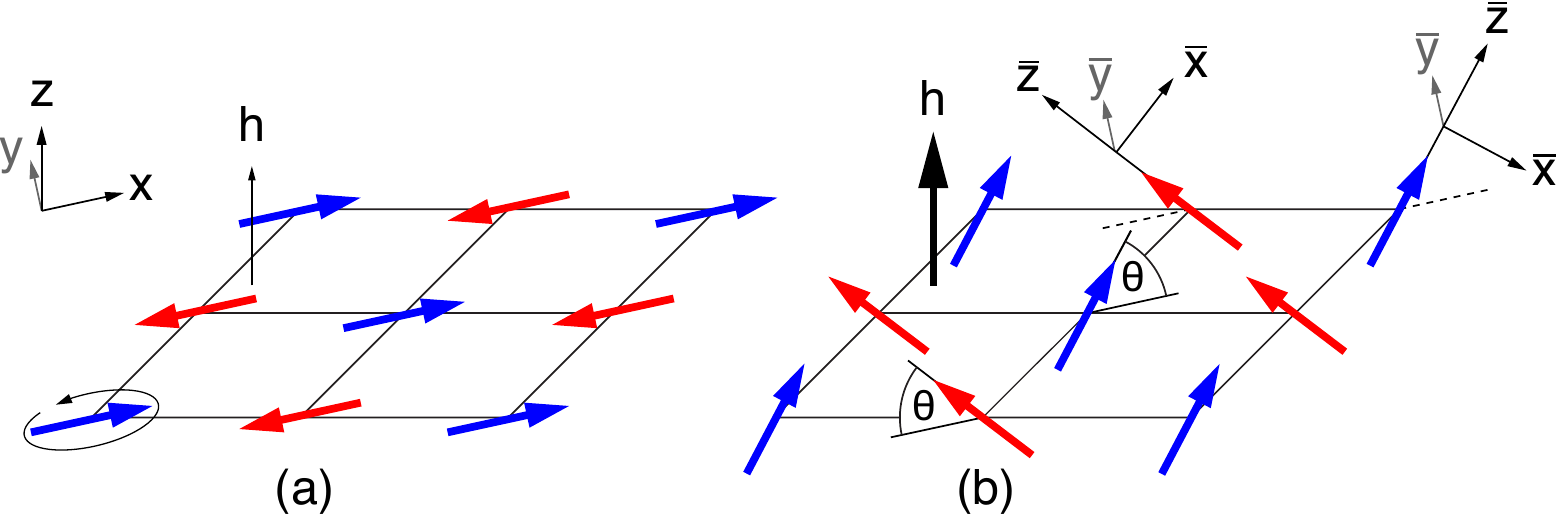}}
\caption{
\emph{(Color online) Classical Heisenberg model in a magnetic field $h$ directed along the $z$-axis. (a) For small fields, the spins are aligned antiferromagnetically in the $xy$-plane. The $O(3)$ symmetry of the model without magnetic field is reduced to $O(2)$ rotations within this plane. (b) For stronger fields, the spins develop a uniform component along the field and are thus canted out of plane. At the saturation field $h_s=8JS$, the spins are aligned ferromagnetically.} 
\label{fig:classical}}
\end{figure}

The aim of the present work is to provide a comprehensive numerical study of the Heisenberg antiferromagnet in magnetic fields ranging from zero up to saturation. Rather than focussing on a particular feature, we aim to give an overview of the various properties induced by the coupling to the magnetic field. Since our results for the dynamical spin structure factors are directly comparable to the inelastic neutron scattering cross section, we believe that this study is of interest to both theorists and experimentalists. 

The paper is organized as follows: In a first part, we discuss static properties of the Heisenberg model in a magnetic field, presenting results for the field dependence of the transverse susceptibility $\chi_\perp$ (Sect.~\ref{sec:chi}), the spin-stiffness $\rho_s$ (Sect.~\ref{sec:rhos}), the spin-wave velocity $c$ (Sect.~\ref{sec:c}), the static spin structure factors (Sect.~\ref{sec:sf}), and the staggered moment $m_s$ (Sect.~\ref{sec:ms}). In Sect.~\ref{sec:dynamics}, the main part of this work, we present our results for the dynamical spin correlation functions obtained by exact diagonalizations of clusters with up to 64 sites. Our conclusion can be found in Sect.~\ref{sec:conclusion}. The reader primarily interested in the dynamical spin structure factors can skip the static part and focus on Sect.~\ref{sec:dynamics}.


\section{Model}
The isotropic spin-$\frac{1}{2}$ Heisenberg model  on the square lattice with nearest-neighbor interactions is described by the Hamiltonian
\begin{equation} \label{eq:hamiltonian}
H=J \sum_{\left\langle i,j\right\rangle} {\bf S}_i \cdot {\bf S}_j - g \mu_B h \sum_i S_i^z \ ,
\end{equation}
where ${\bf S}_i$ is a spin-$\frac{1}{2}$ operator on site $i$, the summation $\left\langle i,j\right\rangle$ extends over nearest-neighbor pairs, and $h$ is a constant magnetic field applied along the quantization axis. In what follows, we set $g \mu_B\to1$, $g$ being the gyromagnetic ratio and $\mu_B$ the Bohr magneton. If not stated explicitly, energies are measured in units of the antiferromagnetic exchange coupling $J>0$. For a system with $N$ spins, the magnetization of a given state is defined as $m=\sum_i \left\langle S_i^z \right\rangle /N$. The fully polarized limit is thus reached at $m=1/2$.

In the absence of a magnetic field, the ground state of the classical model is the N\'eel state, in which all spins are aligned antiparallel with respect to one another, spontaneously breaking the $O(3)$ symmetry of the Hamiltonian [Eq.~(\ref{eq:hamiltonian})]. As the magnetic field is turned on, the symmetry of the model is reduced to $O(2)$, describing rotations around the direction of the applied field. For small fields, the spins preferably align antiferromagnetically in the plane perpendicular to the magnetic field, with a small uniform out-of-plane component, as depicted in Fig.~\ref{fig:classical}. This uniform component becomes stronger with increasing field, until at the saturation field $h_s=8JS$ all spins are aligned ferromagnetically. In the canted regime, the orientation of the spins can be decomposed into a staggered part perpendicular to the magnetic field and a uniform component directed along the field.

The low-energy long-wavelength properties of the Heisenberg model are well described by a nonlinear $\sigma$ model~\cite{chakravarty89,fisher89} whose Lagrangian density is defined as, 
\begin{equation} \label{eq:lagrangian}
{\mathcal L} = -\frac{\rho_s}{2} \left(\nabla {\bf n}\right)^2+\frac{\chi_\perp}{2}\left({\dot {\bf n}}-{\bf h} \times {\bf n}\right)^2 \ .
\end{equation}
Here ${\bf n}$ is a three-dimensional vector representing the orientation of the staggered spin component subject to the constraint ${\bf n}^2=1$, ${\bf h}$ is a constant magnetic field, and a dot denotes the time derivate. This model has three physical parameters: the spin-stiffness $\rho_s$, the staggered moment $m_s$ and the uniform magnetic susceptibility $\chi_\perp$ in the direction perpendicular to ${\bf n}$. The spin-wave velocity $c$ is obtained from the hydrodynamic relation~\cite{halperin69} \begin{equation}\label{eq:hydrodynamic}
c^2=\rho_s/\chi_\perp \ .
\end{equation}
The $\sigma$ model description is valid at small fields, where the staggered moment $m_s$ is large. In addition to being renormalized by quantum fluctuations, see, e.~g., Refs.~\onlinecite{igarashi92,sandvik97}, these microscopic parameters also depend significantly on the strength of the magnetic field. The spin-wave velocity for instance decreases with $h$ and vanishes at the saturation field $h_s$. We discuss the field dependence of these three parameters in Sec.~\ref{sec:static}.

\begin{figure}
\centerline{\includegraphics[width=0.95\linewidth,clip]{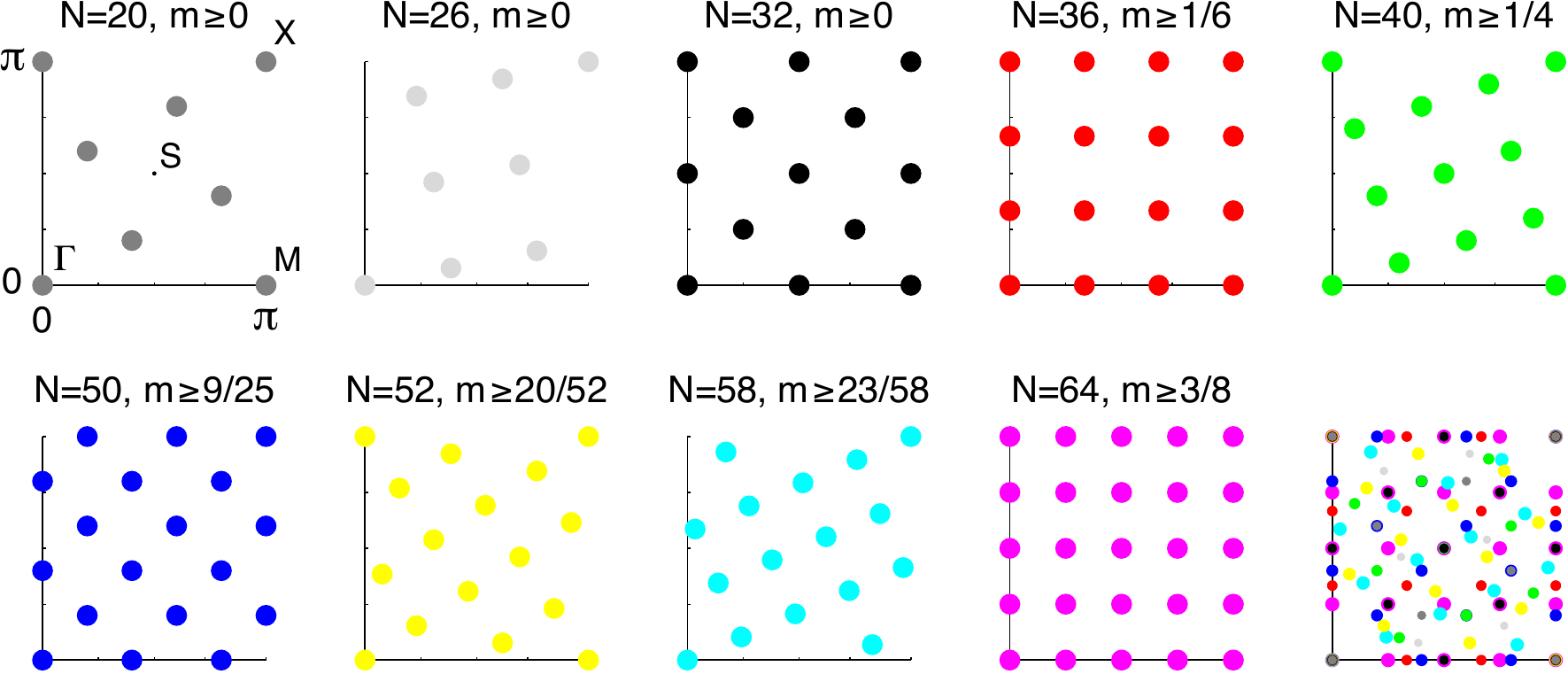}}
\caption{
\emph{(Color online) Reciprocal lattices of the different finite-size clusters used in this work. Zero-field results are obtained from the 32-sites sample, while for the polarized regime, exact diagonalizations of clusters with up to $N=64$ sites can be performed. The achievable magnetizations are indicated for every cluster. Saturation is reached at $m=1/2$.}
\label{fig:momenta}}
\end{figure}

Finite-size effects in an $O(n)$ $\sigma$ model have been discussed in Refs.~\onlinecite{neuberger89,hasenfratz90,hasenfratz93}. For a quadratic sample with $N$ sites, the ground state energy density scales as
\begin{multline}
\label{eq:scalingc} 
e_0(N) = e_0 - 1.437745 \left(\frac{n-1}{2}\right) \frac{c}{N^{3/2}} \\Ê+ 
\frac{(n-1)(n-2)}{8} \frac{c^2}{\rho_s N^2}+{\mathcal O}\left(\frac{1}{N^{5/2}}\right) \ ,
\end{multline}
allowing one to extract the spin-wave velocity $c$. Note that for $n=2$, there is no $1/N^2$ contribution. The staggered magnetization behaves as~\cite{neuberger89,sandvik97}
\begin{equation}\label{eq:scalingms}
m_s^2(N) = m_s^2 \left[1+0.62075\left(\frac{n-1}{2}\right) \frac{c}{\rho_s \sqrt{N}} + {\mathcal O}\left(\frac{1}{N}\right) \right] \ .
\end{equation}
We will use these extrapolation formulas in Sec.~\ref{sec:static} to estimate the spin-wave velocity and the staggered magnetization in the thermodynamic limit, taking into account the reduction of the symmetry from $n=3$ to $2$ in the presence of a magnetic field.


\section{Static properties\label{sec:static}}
In this Sect., we discuss static properties of the Heisenberg model. Readers more interested in the dynamical spin structure factors can directly proceed to Sect.~\ref{sec:dynamics}. Our results are obtained by means of exact diagonalizations and quantum Monte-Carlo simulations using the SSE application~\cite{sandvik99,alet05} of the ALPS library~\cite{alps07}. The reciprocal lattices of the different clusters used in this work are shown in Fig.~\ref{fig:momenta}, together with the accessible magnetizations. Throughout this paper, we use the same color code for all finite-size data. Taking into account translation symmetries, the Hilbert space for a cluster of $N$ sites at magnetization $m$ encompasses roughly ${N \choose N/2-mN}/N$ states. Computations at low fields are most demanding, because of the huge Hilbert spaces requiring a significant amount of main memory. At zero field, exact diagonalizations in the present study are limited to 32 or 36 sites, while in the high-field regime, we can include data from systems with up to 64 sites. The largest Hilbert spaces involved encompass up to several 100 million states. Because the magnetization $m$ is a conserved quantum number, it is computationally advantageous to work with fixed $m$ rather than at a given magnetic field $h$, as it is the case in experiments or in spin-wave calculations. To avoid inexact transformations between these two conjugate variables, we have chosen to present all our numerical results as a function of the magnetization. Comparisons with spin-wave calculations are established by mapping the magnetic field $h$ onto $m$ via the inverse magnetization curve $h(m)$ obtained within linear spin-wave theory. This choice is motivated by the fact that we prefer to focus on exact numerical results, keeping spin-wave approximations at the simplest level that qualitatively captures the physical properties.
\begin{figure}
\centerline{\includegraphics[width=0.95\linewidth,clip]{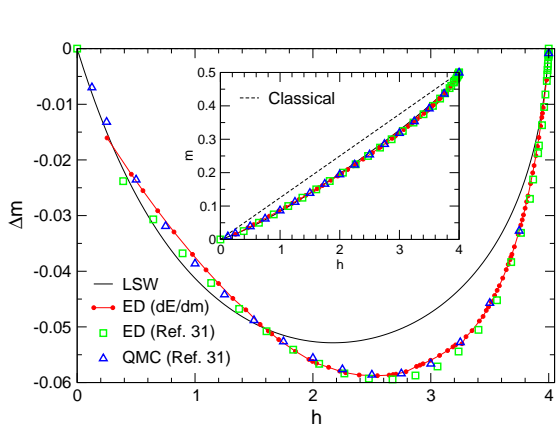}}
\caption{
\emph{(Color online) Inset: Uniform magnetization $m$ as a function of the applied magnetic field $h$. The dashed line is the classical result. Main panel: Difference to the classical curve $\Delta m(h)=m(h)-h/8$. The solid black line represents the linear spin-wave approximation. Small red dots are obtained from the derivative of the extrapolated ground state energy, open squares represent exact diagonalization results from clusters with at least 40 sites, and open triangles indicate quantum Monte-Carlo results at temperatures $T\le 0.02J$.}
\label{fig:m}}
\end{figure}

\subsection{Uniform magnetization\label{sec:m}}
On finite-size samples, the magnetization process occurs in finite steps. To obtain a continuous magnetization curve, one can either interpolate these steps along $h$ and $m$ or invert the derivative of the extrapolated finite-size ground state energy [Eq.~(\ref{eq:scalingc})] with respect to the magnetization $\partial e_0/ \partial m$. Using the finite-size scaling has the disadvantage that the energies obtained from exact diagonalizations first have to be interpolated with respect to the magnetization $m$ before they can be extrapolated to the thermodynamic limit because the magnetization steps of different samples only coincide for a few particular values of $m$. The interpolation is unproblematic, because the energy is a smooth function of the magnetization. Subsequently, the extrapolation is performed for all values of $m$ that are realized in at least one sample. The magnetization curve of the Heisenberg model has been calculated before by exact diagonalizations of clusters containing at least 40 sites~\cite{richter04} and also quantum Monte-Carlo simulations~\cite{sandvik99b,richter04}. We have reproduced these results in Fig.~\ref{fig:m}, together with our values obtained from the derivation of the energy. The actual magnetization curve is presented in the inset, while the main panel shows the difference to the linear classical behavior. 
The overall agreement between linear spin-wave approximation~\cite{zhitomirsky98} and numerical results is quite remarkable, the difference being everywhere smaller than $5\%$. Our approach of using this magnetization curve to compare spin-wave results with numerical simulations is thus justified. The magnetization curve obtained from the extrapolated ground state energy is in excellent agreement with Monte-Carlo simulations, especially at higher fields, where larger samples can be used for the extrapolation. Close to saturation, the magnetization is expected to behave linearly with logarithmic corrections~\cite{zhitomirsky98,gluzman93,sachdev94,honecker99}, in contrast to the root singularities encountered in one-dimensional systems~\cite{dzhaparidze78,honecker99}.

\subsection{Transverse susceptibility\label{sec:chi}}
\begin{figure}
\centerline{\includegraphics[width=0.95\linewidth,clip]{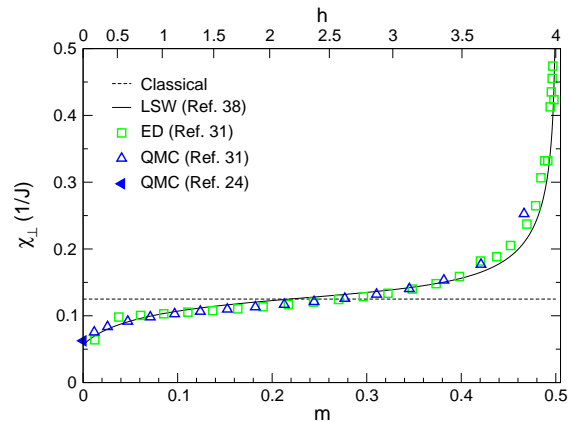}}
\caption{
\emph{(Color online) Transverse susceptibility $\chi_\perp$ as a function of the magnetization $m$ and the applied magnetic field $h$. The classical result (dashed line) is constant, while the linear spin-wave curve (solid line) obtained as the numerical derivative of the uniform magnetization diverges in the limit $m \to 1/2$ because of the logarithmic singularity in $m(h)$.}
\label{fig:chi}}
\end{figure}
The magnetic susceptibility measures the variation of the magnetization with respect to a small external field ${\bf h}'$ applied in addition to the arbitrarily large field $h$ introduced in Eq.~(\ref{eq:hamiltonian}). We can thus construct a susceptibility tensor as
\begin{equation*}
\chi^{\alpha\beta} = \left. \frac{\partial \left\langle 0 \right|ÊS^\alpha \left| 0 \right\rangle}{\partial h'^\beta} \right| _{h'^\beta\to0} \ .
\end{equation*}
Defining the terms parallel and perpendicular with respect to the direction of the staggered moment, the transverse susceptibility $\chi_\perp$ measures the variation of the uniform magnetization with respect to the applied field
\begin{equation*}
\chi_\perp = \frac{\partial m}{\partial h} \ .
\end{equation*}

\begin{figure}
\centerline{\includegraphics[width=0.95\linewidth,clip]{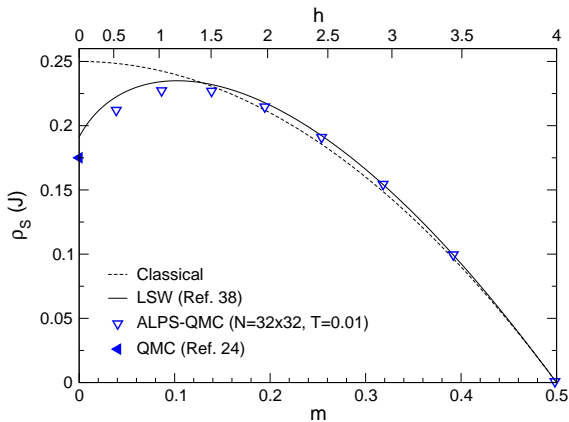}}
\caption{\emph{Spin-stiffness $\rho_s$ as a function of the magnetization $m$ and the applied magnetic field $h$. The results have been obtained by quantum Monte-Carlo simulations~\cite{sandvik99,alet05} using the ALPS~\cite{alps07} application and spin-wave calculations. The initial increase of the spin-stiffness can be attributed to a reduction of quantum fluctuations, while the linear decrease at high fields is due to the canting of the spins towards the direction of the applied field.}
\label{fig:rhos}}
\end{figure}

\begin{figure}
\centerline{\includegraphics[width=0.95\linewidth,clip]{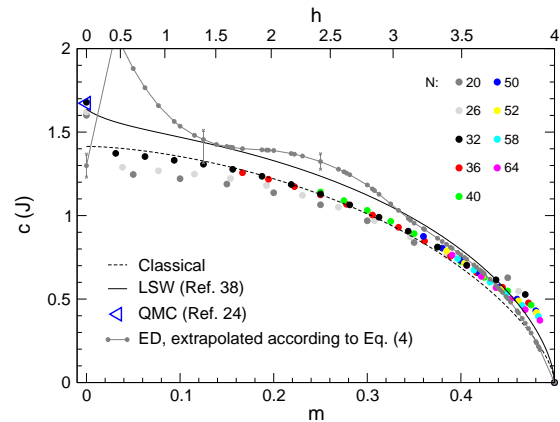}}
\caption{
\emph{(Color online) Spin-wave velocity $c$ as a function of the magnetization $m$ and the applied field $h$. Symbols correspond to the slope of the dispersion extracted from finite-size clusters of size $N$. Small gray dots represent the thermodynamic values obtained by extrapolations according to Eq.~(\ref{eq:scalingc}). The dashed and the solid lines are spin-wave results.}
\label{fig:c}}
\end{figure}

In Fig.~\ref{fig:chi}, we plot the transverse susceptibility as a function of the magnetization and the magnetic field. The upper $x$-axes shown in Figs.~\ref{fig:chi}-\ref{fig:c} and \ref{fig:ms} represent the linear spin-wave mapping between the magnetization $m$ and the magnetic field $h$, as a function of which spin-wave results are unambiguous. \emph{Note that the classical results are plotted using the classical magnetization curve.} The zero-field quantum Monte-Carlo result is taken from Ref.~\onlinecite{sandvik97}, which we have arbitrarily chosen as the representative of a large body of works encompassing many different methods. The spin-wave result has been derived in Ref.~\onlinecite{chernyshev09}. Exact diagonalization and quantum Monte-Carlo data~\cite{richter04} at finite field are obtained by numerically differentiating the magnetization curves shown in Fig.~\ref{fig:m}. In the classical limit, the susceptibility is constant, $\chi_\perp^0 = 1/8$, and the hydrodynamic relation~[Eq.~(\ref{eq:hydrodynamic})] is exactly satisfied for all magnetizations. Quantum fluctuations modify the shape of $\chi_\perp$ considerably and lead to the divergence in the limit $m\to1/2$, stemming from the logarithmic singularity in the magnetization curve. We note that the linear spin-wave approximation captures these effects extremely well and is in perfect agreement with Monte-Carlo data.

\subsection{Spin-stiffness \label{sec:rhos}}
The elastic energy required to twist a spin arrangement is proportional to the spin-stiffness $\rho_s$, see Eq. (\ref{eq:lagrangian}). For a planar order parameter of the form
\begin{equation*}
{\bf n}({\bf r}) = \left[\cos \varphi({\bf r}), \sin \varphi({\bf r}),0\right] \ ,
\end{equation*}
that slowly rotates with a constant twist $\delta\varphi=\varphi({\bf r}_i)-\varphi({\bf r}_j)$ between adjacent sites ${\bf r}_i$ and ${\bf r}_j$, the difference between the energy density of the twisted and the collinear spin arrangement is given by 
\begin{equation*}
\delta e = e_0(\delta \varphi) - e_0(\delta \varphi = 0) \approx J \rho_s \left(\delta\varphi\right)^2 \ ,
\end{equation*}
where the twist is applied along both space directions. In principle, it is possible to calculate $\rho_s$ in exact diagonalizations~\cite{einarsson95}, but it is far easier and more accurate to obtain the spin-stiffness from quantum Monte-Carlo simulations as the global winding number fluctuations~\cite{pollock84}. In Fig.~\ref{fig:rhos}, we plot $\rho_s$ as a function of the magnetization and the magnetic field. Previous results at zero field are taken from Ref.~\onlinecite{sandvik97}. Given the simplicity of the linear spin-wave approximation, the excellent agreement with Monte-Carlo results is remarkable. Because the spin-stiffness is proportional to the elastic energy required to deform the collinear N\'eel order, it provides a measure of the ordering tendencies, or inversely, the quantum fluctuations present in the antiferromagnet. The initial increase of $\rho_s$ with the magnetization can be attributed to a reduction of quantum fluctuations. The almost linear decrease at high fields however is due to the canting of the spins and has nothing to do with fluctuations because the exact result is almost identical to the classical curve. The reduction of quantum fluctuations at low fields can also be seen in the staggered magnetization discussed in Sect.~\ref{sec:ms}.

\subsection{Spin-wave velocity\label{sec:c}}
\begin{figure}
\centerline{\includegraphics[width=0.95\linewidth,clip]{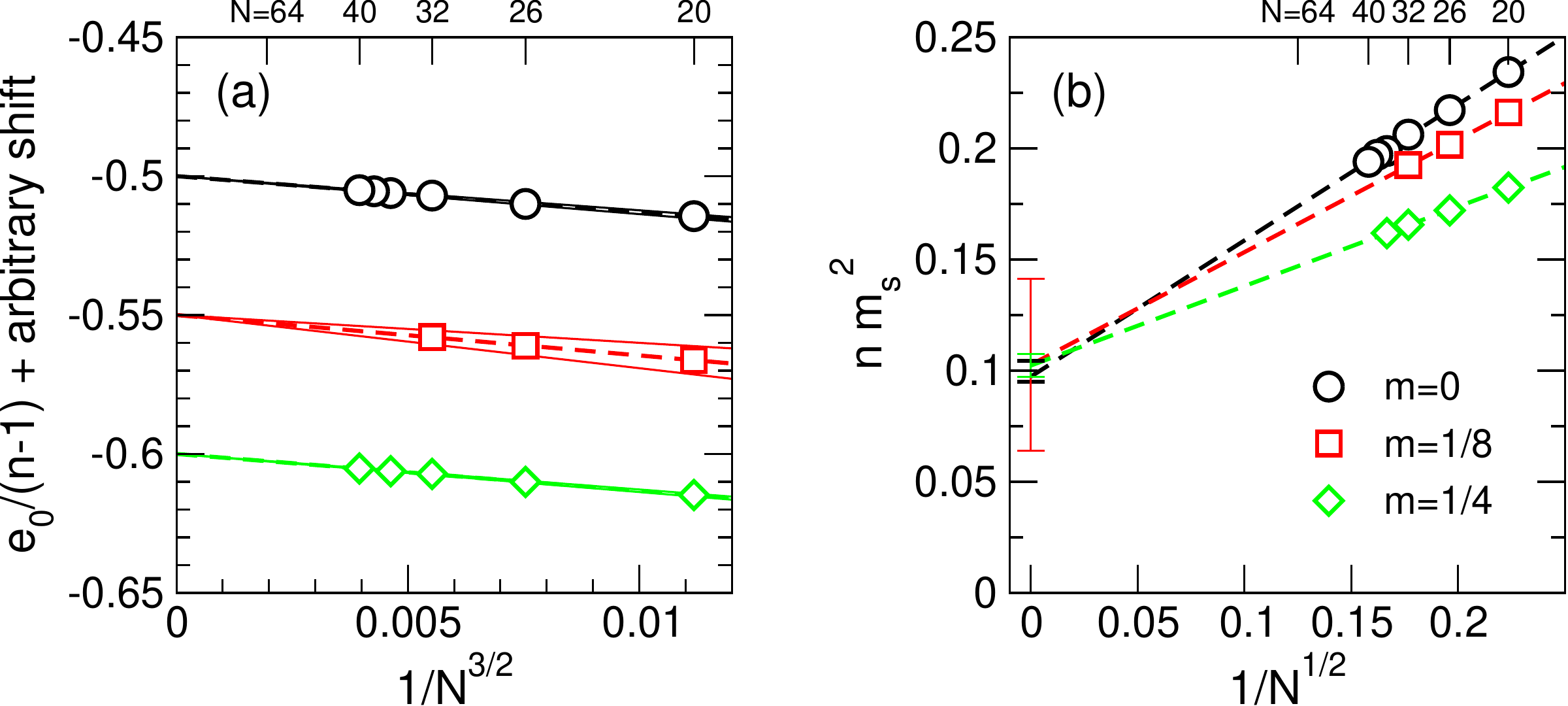}}
\caption{
\emph{(Color online) (a) Finite-size extrapolations (dashed lines) of the energy $e_0$ according to Eq.~(\ref{eq:scalingc}) for magnetizations $m=0$, $1/8$, and $1/4$. The spin-wave velocity $c$ is equal to the slope of the extrapolations. The error bars shown in Fig.~\ref{fig:c} correspond to the $95\%$ confidence intervals of the above fits, indicated by the thin solid lines. (b) Finite-size extrapolations of the staggered moment $m_s^2$ according to Eq.~(\ref{eq:scalingms}). The error bars represent the $95\%$ confidence levels of the linear regressions, also shown in Fig.~\ref{fig:ms}. Zero-field data from clusters with more than 32 sites are taken from Ref.~\onlinecite{richter04}.
 }
\label{fig:extrapolations}}
\end{figure}
There are two ways to extract the spin-wave velocity $c$ from exact diagonalization results: One can either determine the slope of the dispersion in the vicinity of the ground state momentum, or, one can fit the finite-size scaling of the ground state energy to Eq.~(\ref{eq:scalingc}) and determine $c$ from the fitting parameters. This latter method has the disadvantage that the energies first have to be interpolated with respect to the magnetization $m$, before the extrapolation can be performed, as discussed in Sect.~\ref{sec:m}. In Fig.~\ref{fig:c}, we show the results from both approaches and compare them to spin-wave calculations~\cite{chernyshev09}. 

While the three methods give qualitatively similar results, the quantitative differences are non-negligible, especially in the low-field regime. Surprisingly, the zero-field spin-wave velocity extracted from the sample with 32 sites is in perfect agreement with quantum Monte-Carlo data~\cite{sandvik97}. The results from the 20 and 26-sites cluster are somewhat lower, suggesting a thermodynamic value slightly above the one obtained from the 32-sites system. The abrupt drop of the velocity upon turning on the magnetic field is unphysical and the pronounced finite-size effects observed in this regime indicate that the actual value  of $c$ is substantially larger than the 32-sites cluster result. For increasing field, these effects become smaller and one could conclude that at $m=0.2$, the thermodynamic value should be very close to the 36-sites result. This is however not quite correct because for any system size, the distance to the gapless mode is at least $\pi/4$, see Fig.~\ref{fig:momenta} for the location of the momenta in different clusters, and thus still far from the cone center. For small and intermediate fields, one therefore systematically underestimates the spin-wave velocity. At very high fields, the direction of the finite-size effects is reversed because the dispersion starts to resemble the ferromagnetic $\epsilon_{\bf k} \propto k^2$ form and the finite-size momenta are too far away from the gapless mode to capture this behavior. This is best seen in the longitudinal structure factors shown in Fig.~\ref{fig:longall}. 

\begin{figure*}[t]
\centerline{\includegraphics[width=0.95\linewidth,clip]{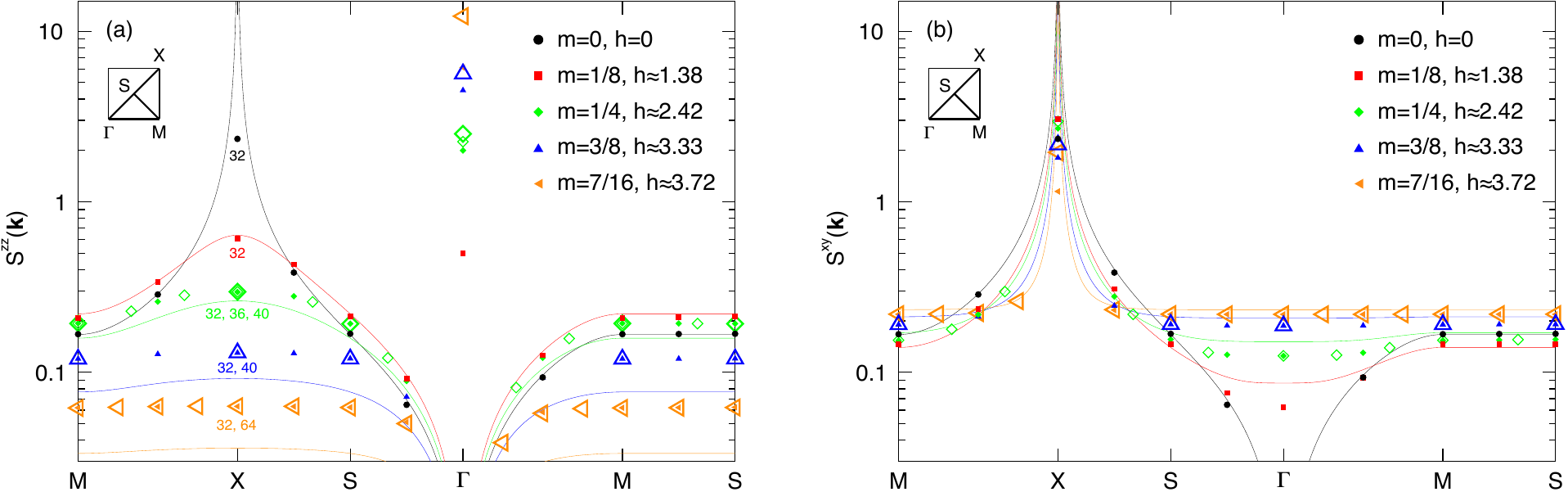}}
\caption{\emph{(Color online) Static longitudinal (a) and transverse (b) spin structure factors $S^{\alpha\beta}({\bf q})$ for various magnetizations $m$ along a path connecting highly symmetric points in the Brillouin zone. Symbols represent exact diagonalization results, with numbers indicating the system sizes. Larger symbols stand for bigger clusters.  Solid lines show the spin-wave results derived in Ref.~\onlinecite{zhitomirsky99}.}
\label{fig:sf}}
\end{figure*}

Given the pronounced finite-size effects at small fields, one can argue that an extrapolation of the ground state energies according to Eq.~(\ref{eq:scalingc}) would give a more accurate value of $c$. On the contrary, at zero field, this method clearly underestimates the spin-wave velocity, as already noted in Ref.~\onlinecite{schulz96}, while for small $m$, it largely overestimates $c$. We think that the huge initial increase is due to the change of symmetry: While the extrapolation formula is modified when turning on the magnetic field, from $n=3$ to $2$, see Eq.~(\ref{eq:scalingc}), the finite-size clusters are too small to reflect this change. This interpretation is consistent with the observed overestimation of almost a factor $3/2$. A reason for the underestimation of the zero-field spin-wave velocity might be the influence of higher order terms in the fitting formula [Eq.~(\ref{eq:scalingc})], which for $n=3$ also has a contribution $1/N^2$. Including higher order corrections leads to an increase of the velocity $c$, but also to inconclusively large error bars. We believe that the bump observed between $m=0.2$ and 0.3 is an artifact of the finite-size extrapolations, as it seems driven by the $36$ and $40$-sites cluster. To give an idea of the variability of the extrapolations, the fits for $m=0$, $1/8$, and $1/4$ are shown in Fig.~\ref{fig:extrapolations}(a). The $95\%$ confidence intervals are shown as thin solid lines. They are also indicated in Fig.~\ref{fig:c}. Concerning more advanced spin-wave calculations of the dispersion, we would like to refer the reader to a recent work by Kreisel {\it et al.}~\cite{kreisel08}, discussing a non-analytic behavior of the spin-wave velocity induced by the magnetic field. The validity of the hydrodynamic relation in a magnetic field has been confirmed very recently in Ref.~\onlinecite{chernyshev09} within a spin-wave approach. At zero field, Hamer {\it et al.}~\cite{hamer94} verified this property using series expansions and spin-wave approximations.

\subsection{Static spin structure factor\label{sec:sf}}
The static spin structure factor
\begin{equation*}
S^{\alpha\beta}({\bf k}) = 
\frac{1}{N} \sum_{i,j} e^{i {\bf k} \cdot \left({\bf r}_i-{\bf r}_j\right)}
 \left\langle S_i^\alpha S_j^\beta\right\rangle \ ,
\end{equation*}
with $\alpha,\beta=x,y,z,+,-$, is the Fourier transformation of the equal-time spin-spin correlators and can be directly measured in elastic neutron scattering. Fig.~\ref{fig:sf} shows the longitudinal and transverse structure factors obtained from exact diagonalizations, see also Ref.~\onlinecite{yang97}, and spin-wave calculations~\cite{zhitomirsky99} for several magnetizations along highly symmetric points in the Brillouin zone. Because there is no spontaneous symmetry breaking in finite systems, the exact diagonalization results for $S^{xx}({\bf q})$ and $S^{yy}({\bf q})$ are identical. Hence, it is useful to define a transverse component $S^{xy}$ as
\begin{equation*}
S^{xy}({\bf k}) =\frac{1}{4} \left[S^{+-}({\bf k})+S^{-+}({\bf k})\right] \ .
\end{equation*}
In contrast, the spin-wave results are derived in the symmetry broken phase where the staggered spin component is directed along the $x$-axis. The transverse correlations shown in Fig.~\ref{fig:sf}(b) represent the average $2S^{xy}({\bf k}) =S^{xx}({\bf k})+S^{yy}({\bf k})$. At zero field, the structure factors reflect the antiferromagnetic order with a pronounced peak at $X=(\pi,\pi)$. As the magnetic field is turned on, this contribution rapidly decreases in the longitudinal channel but remains dominant in the transverse structure factor, indicating staggered order in the plane perpendicular to the magnetic field. With increasing field, the spins are gradually canted out of the $xy$-plane and develop a uniform component that manifests itself at momentum $\Gamma=(0,0)$ in the longitudinal structure factor. At high fields, both curves are almost flat, except at the before mentioned two points. We note that linear spin-wave results capture the essential features qualitatively correctly, apart from the uniform contribution to the longitudinal structure factor. Our results are in good agreement with recent quantum Monte-Carlo data~\cite{syljuasen08}.

\subsection{Staggered magnetization\label{sec:ms}}
\begin{figure}
\centerline{\includegraphics[width=0.95\linewidth,clip]{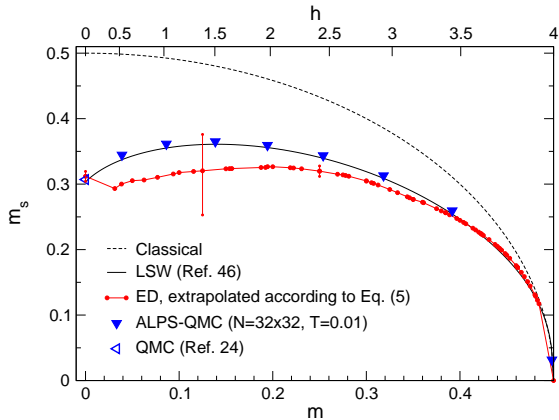}}
\caption{
\emph{(Color online) Staggered magnetization $m_s$ obtained from finite-size extrapolation according to Eq.~(\ref{eq:scalingms}), quantum Monte-Carlo simulations, and spin-wave calculations.}
\label{fig:ms}}
\end{figure}

Similar to the uniform magnetization, the staggered moment $m_s$ can be obtained as the derivative of the ground state energy density with respect to the staggered field. The linear spin-wave result has been derived in Ref.~\onlinecite{spremo06}. Alternatively, $m_s$ can be obtained from the transverse structure factor as
\begin{equation*}
m_s^2 = \frac{1}{N'} S^{xy}\left({\bf k}={\bf Q}\right) \ , \quad N'=\begin{cases} N+2 & (h=0) \\ N-1 &(h>0)\end{cases}
\end{equation*}
where $N'$ is a normalization factor. On finite systems of size $N$, the static structure factors satisfy the relation
\begin{equation*}
S^{+-}({\bf k}) = S^{-+}({\bf k}) + 2 m \ ,
\end{equation*}
contributing a term $1/N'$ to $m_s^2$. This additional term makes the extrapolation according to Eq.~(\ref{eq:scalingms}) more difficult because one needs to take into account an additional parameter. To circumvent this problem, we only consider the $S^{-+}({\bf k})$ contribution to the transverse structure factor. In the thermodynamic limit, both choices of the normalization $N'$ are identical, but on the small samples we study, there are sizable differences. The zero-field choice was introduced in Ref.~\onlinecite{bernu92} and successfully applied to the $J_1$-$J_2$ model on the square lattice~\cite{schulz96}. In the presence of a magnetic field, using only the $S^{-+}({\bf k})$ contribution which vanishes at ${\bf k}=\Gamma$ naturally leads to a normalization $N'=N-1$. In Fig.~\ref{fig:ms}, we show the staggered magnetization obtained from extrapolations based on Eq.~(\ref{eq:scalingms}), quantum Monte-Carlo simulations and spin-wave calculations.
At zero field, quantum fluctuations are strong, leading to a considerable reduction of the ordered moment. In this case, all methods give a similar value slightly above $m_s \approx 0.3$. As the field is turned on, $m_s$ increases and the fluctuations thus become less important, giving rise to an almost classical regime, in which the dispersion is well described by linear spin-wave theory. This aspect will be discussed in more detail in Sect.~\ref{sec:dynamics}. After the maximum at $m\approx 0.15$, the staggered moment is again reduced because of the canting of the spins and eventually vanishes in the saturation limit. While the extrapolation of exact diagonalization data according to Eq.~(\ref{eq:scalingms}) gives a very good result at  zero field~\cite{schulz96,richter04}, it considerably underestimates the staggered moment at small fields. To get an idea of the variability of the extrapolations, fits at magnetizations $m=0$, $1/8$, and $1/4$ are shown in Fig.~\ref{fig:extrapolations}. The $95\%$-confidence intervals are shown as thin solid lines. They are also indicated in Fig.~\ref{fig:ms}. These confidence intervals should not be interpreted in a strict statistical sense, but are rather intended to remind the reader of the fact that the extrapolations to the thermodynamic limit are based on only a handful of small clusters. The inaccuracy at small fields is related to the smallness of the clusters used for the extrapolations as well as the lack of analytical predictions for higher order finite-size corrections. Similar to the extraction of the spin-wave velocity, the method is much more accurate at higher fields, where bigger samples can be taken into account. In contrast to finite-size extrapolations, spin-wave results are in excellent agreement with quantum Monte-Carlo data obtained on a cluster with $N=32\times32$ sites.


\section{Dynamical properties\label{sec:dynamics}}
In the main part of this paper, we study the dynamical spin correlation functions
\begin{equation} \label{eq:dynsf}
S^{\alpha\beta}(\omega,{\bf q}) = \frac{1}{N} \sum_{i,j} e^{i {\bf q} \cdot \left({\bf r}_i-{\bf r}_j\right)}
\int_{-\infty}^\infty \!\! dt \, e^{i \omega t} \left\langle S_i^\alpha(t) S_j^\beta(0)\right\rangle \ ,
\end{equation}
where, as before, $N$ is the number of sites on the lattice. 
This quantity is directly measurable in inelastic neutron scattering experiments. In what follows, we consider $\alpha\beta=zz$, which we refer to as longitudinal correlations because they are parallel to the applied field, and $\alpha\beta=+-, -+$, allowing us to construct the transverse correlation functions defined as
\begin{equation*}
S^{xy}(\omega,{\bf q}) = \frac{1}{4}\left[S^{+-}(\omega,{\bf q})+S^{-+}(\omega,{\bf q})\right] \ .
\end{equation*}
\begin{figure*}
\centerline{\includegraphics[width=0.95\linewidth,clip]{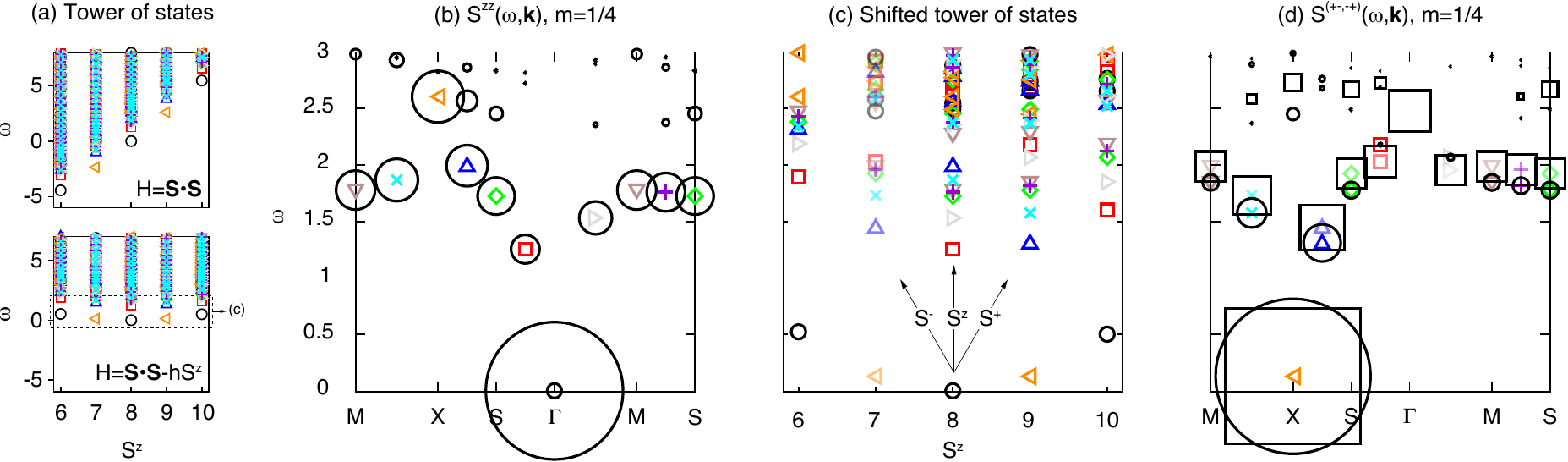}}
\caption{
\emph{(Color online) 
Tower of states (a,c) and longitudinal (b) and transverse (d) structure factors obtained from the 32-sites cluster at magnetization $m=1/4$, corresponding to $S^z=8$. The ground state energy of the $S^z=8$ levels is set to zero. The location of the poles in the structure factors can be traced backed to the energy levels in the tower of states. To facilitate identification, the strongest signatures are indicated by the symbols of the corresponding eigenlevels. In the presence of a magnetic field, constructing the transverse structure factor is not straightforward because the relevant towers for $S^{+-}$ and $S^{-+}$ start at different energies. The alignment procedure is illustrated in the bottom panel of (a).}
\label{fig:spectrum}}
\end{figure*}

In exact diagonalization one can rather easily calculate spectral functions directly in real frequency, using the so called continued-fraction technique \cite{gagliano87,dagotto94}. Combined with the Lehmann representation
\begin{equation} \label{eq:dynsf2}
S^{\alpha\beta}(\omega,{\bf q}) = \sum_i \left| \left\langle \Psi_i \right| S^\beta({\bf q}) \left| \Psi_m \right\rangle \right|^2 
\delta (\omega-[E_i-E_m]) \ ,
\end{equation}
where $\Psi_m$ is the ground state at a given magnetization $m$ with energy $E_m$ and $\Psi_i$ are excited states with energies $E_i$, this approach allows one to obtain unbiased exact results, including residues, although only on finite-size samples. 

\begin{figure*}
\centerline{\includegraphics[width=0.95\linewidth,clip]{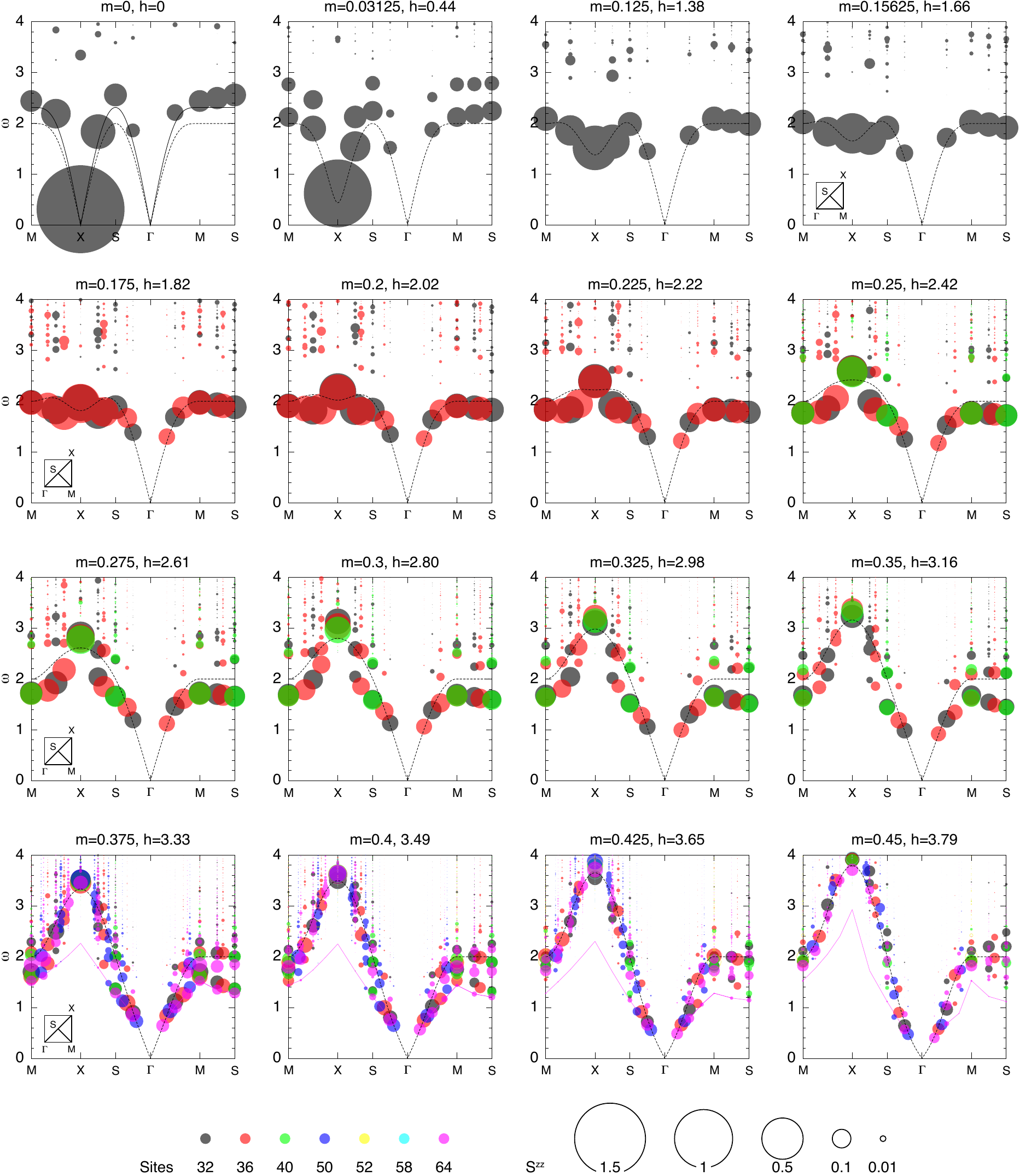}}
\caption{
\emph{Synthetic superposition of the longitudinal dynamical structure factors along a path of highly symmetric points in the Brillouin zone. Different colors represent data from different clusters and the area of the symbols is proportional to $S^{zz}(\omega,{\bf q})$. Dashed lines show the dispersions obtained within linear (harmonic) spin-wave theory and the solid line represents spin-wave results with first-order corrections (only shown for $m=0$). For magnetizations around $m\approx 0.15$, quantum fluctuations are almost negligible and the spin-wave dispersion is in good agreement with numerical results. At higher fields, $m\gtrsim0.3$, fluctuations are again important and lead to the spontaneous decay of magnons. This instability is reflected in a reduction of weight in main peak accompanied by the appearance of small poles at lower energies. This process starts around ${\bf q}=X$ and spreads over almost the whole Brillouin zone.} 
\label{fig:longall}}
\end{figure*}
The Zeeman term $-h \sum_i S^z_i = -h S^z_\text{tot}$ is usually not included in exact diagonalization calculations because it is constant within a given $S^z_\text{tot}$ sector. While this omission does not affect the longitudinal structure factors, extracting the frequency dependence of the transverse spin correlations is more subtle because the two contributions act between different $S^z_\text{tot}$ sectors. This is best understood by tracing back the origin of the poles in the dynamical structure factors to the eigenlevels in the tower of states. In Fig.~\ref{fig:spectrum}(a), we present part of the tower of states for the 32-sites cluster for $6\leq S^z \leq 10$. The top panel shows the spectrum of the Hamiltonian without magnetic field, while in the bottom panel, we have subtracted the contribution $h S^z$, for a magnetic field $h$ corresponding to $m=1/4$, i.e., $S^z=8$ on this cluster. Realigning the tower in the presence of a magnetic field is thus equivalent to shifting the spin lowering and raising contributions to the transverse structure factors in such a way that the lowest-lying poles of the ${\bf k}={\bf Q}$ modes coincide.
The most important properties of the tower of states are the alternation of the ground state momenta between $(0,0)$ and $(\pi,\pi)$ in adjacent $S^z$ sectors and the proportionality of the ground state energies to $S^z(S^z+1)$. 

Figs.~\ref{fig:spectrum}(b) and (d) show the longitudinal and transverse structure factors obtained from this spectrum. In Fig.~\ref{fig:spectrum}(c), we have zoomed in on a part of the tower of states to facilite identification of eigenlevels and poles. As expected, at a magnetization $m=S^z/N$, the main magnon branch of the longitudinal correlations is made of spin $S=S^z$ states. Similarly, in the transverse channel, the magnon dispersion originates from spin $S=S^z\pm1$ states. As an exception to this rule, the pole at $X=(\pi,\pi)$ in the longitudinal structure factor originates from the lowest-lying spin $S=S^z+1$ level, as can be seen in Fig.~\ref{fig:spectrum}(a). Its energy is equal to the magnetic field. In this sequence of pictures, it is easy to see that while the lowest lying poles of the spin raising and lowering contributions to the transverse structure factor coincide, higher lying levels are slightly shifted with respect to one another. 

\subsection{Synthetic view of the dynamical structure factors\label{sec:synthetic}}
In exact diagonalizations, the restriction to relatively small systems makes it difficult to extract wave vector dependent quantities to the thermodynamic limit. Only highly symmetric points like $M=(0,\pi)$ or $S=(\pi/2,\pi/2)$ are common among several systems sizes up to 64 sites. By construction, $\Gamma=(0,0)$ and $X=(\pi,\pi)$ are present in any cluster, but the physics at these points is usually quite simple. In the presence of a magnetic field, bigger clusters can be used for the calculations, but the main difficulty lies in the fact that the magnetization steps are in general not compatible because they also depend on the system size. It is therefore basically impossible to present an exact superposition of all dynamical correlation functions obtained from different systems sizes at a given magnetization. However, since the shape of the dynamical structure factors does not change substantially from one magnetization step to another, it is nevertheless possible to construct quite accurate interpolations. For a given magnetization $m$ and a cluster of size $N$, we chose to either take the exact data if available or instead shift the whole structure factor obtained for $m'$ closest to $m$ to the position determined by linear interpolation between the lowest-lying poles at $m'\pm1/N$. While this interpolation procedure does not yield exact results, it allows one to visualize data from different clusters in a single plot, thus providing a nice overview of the observed features. In Fig.~\ref{fig:longall}, we show such a synthetic superposition of the longitudinal structure factors obtained for various magnetizations along a path through the Brillouin zone. 
The zero-energy mode of the longitudinal structure factor at momentum $(0,0)$ is proportional to the magnetization squared, i.e., $S^{zz}(\omega=0,{\bf q}=\Gamma)=(S^z)^2$. In what follows, we omit this contribution. Note that the magnetization steps in the first row are not equidistant. At low fields, $m<1/6$, we exclusively use the 32-sites cluster for all calculations.
\begin{figure}
\centerline{\includegraphics[width=0.95\linewidth,clip]{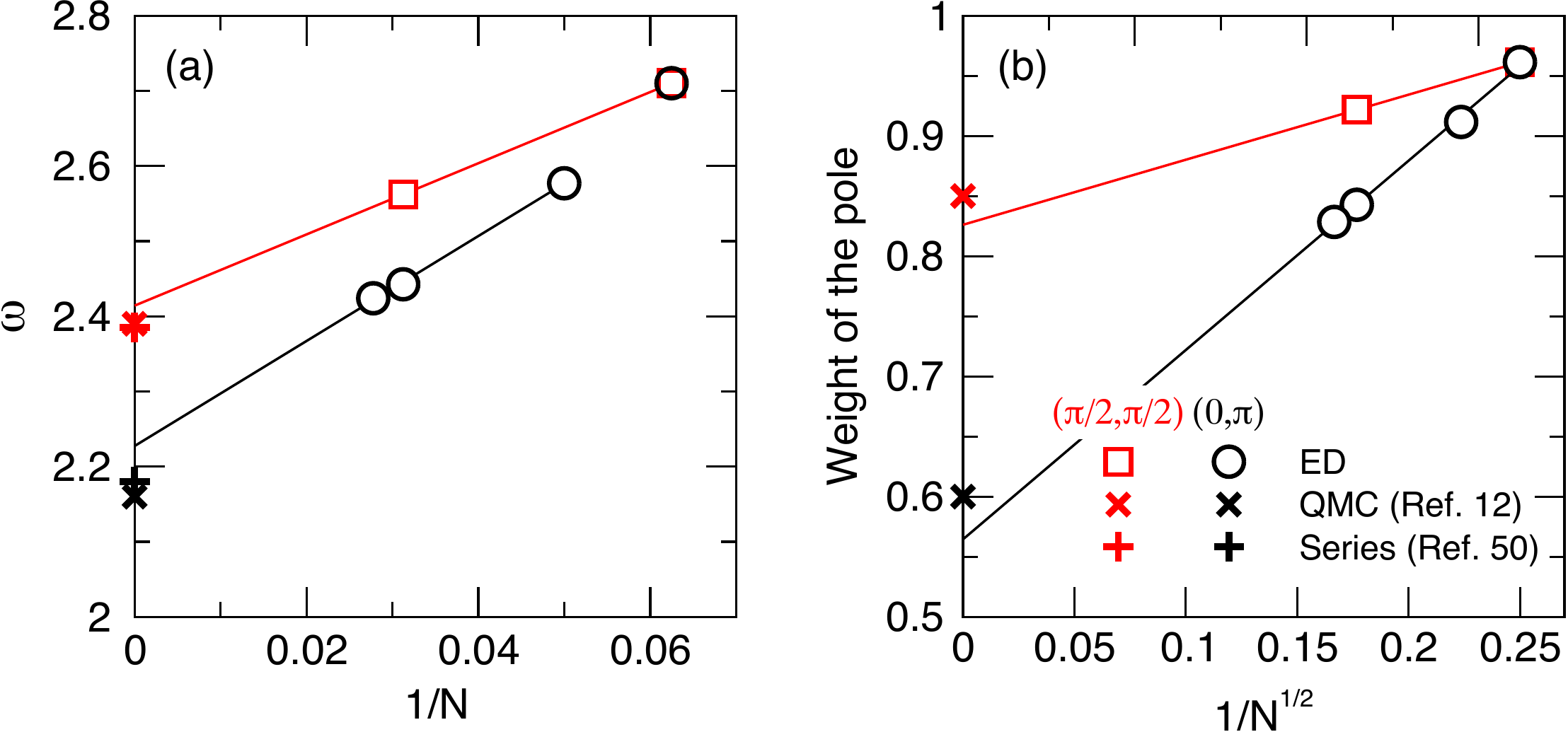}}
\caption{
\emph{Finite-size scaling of the spin-wave energies (a) and the quasi-particle residues (b) at momenta $(\pi/2,\pi/2)$ and $(\pi,0)$.} 
\label{fig:zoneboundary}}
\end{figure}

At zero field, the ground state of the infinite systems spontaneously breaks the $O(3)$ rotational symmetry of the Hamiltonian [Eq.~(\ref{eq:hamiltonian})], leading to the existence of two Goldstone modes. The distinct pole at ${\bf k}=X$ collapses to the ground state energy as $1/N$ because it originates from a spin-1 level that is part of the Anderson tower, see Refs.~\onlinecite{bernu92,anderson52}. The strong intensity reflects the dominantly antiferromagnetic character of the spin alignment. An interesting zero-field property discussed in detail in the literature~\cite{kim99,sandvik01,ronnow01,christensen04,christensen07}, and mentioned in the introduction, is the dispersive feature observed along the magnetic zone boundary between $M$ and $S$. In Fig.~\ref{fig:zoneboundary}, we present a finite-size scaling (solid lines) of the spin-wave energies and the quasi-particle residues at momenta $(\pi/2,\pi/2)$ [$N=16,32$] and $(\pi,0)$ [$N=16,20,32,36$]. The extrapolation to the thermodynamic limit is in surprisingly good agreement with quantum Monte-Carlo simulations~\cite{sandvik01} and series expansion calculations~\cite{zheng05}.
\begin{figure*}
\centerline{\includegraphics[width=0.95\linewidth,clip]{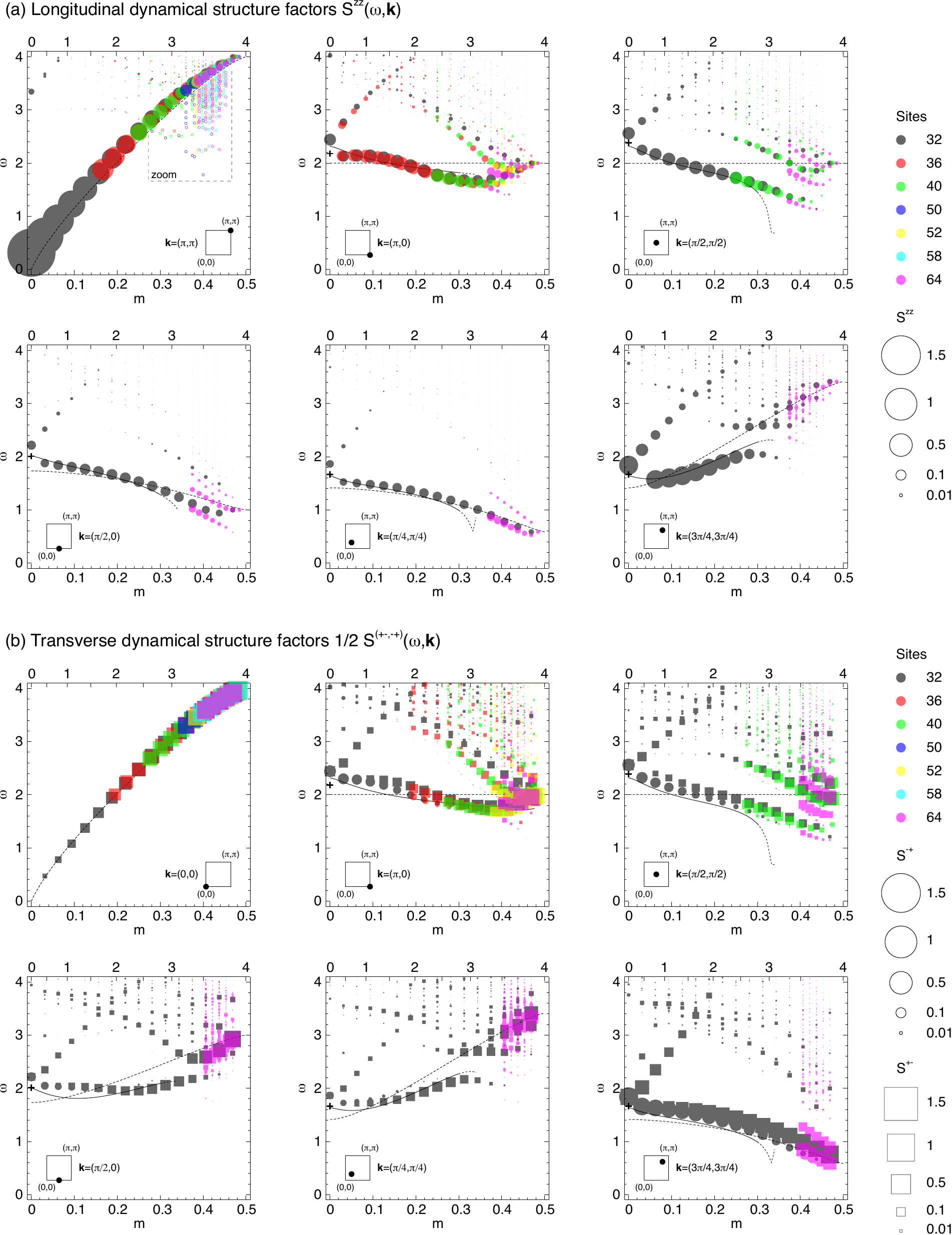}}
\caption{
\emph{Longitudinal (a) and transverse (b) dynamical structure factors at selected momenta. Different colors correspond to different clusters and the area of the symbols is proportional to the correlation functions. Dashed lines represent the dispersion obtained within linear (harmonic) spin-wave theory, while solid lines are spin-wave results with perturbative $1/S$ corrections.The upper $x$-axes indicate the magnetic field obtained within linear spin-wave theory. Crosses represent zero-field series expansion results~\cite{zheng05}. A false-color density plot of the longitudinal spin correlations is shown in Fig.~\ref{fig:dynszz}.} 
\label{fig:dynsf}}
\end{figure*}

Let us now return to the discussion of Fig.~\ref{fig:longall}. A finite magnetic field reduces the symmetry to $O(2)$, which is again spontaneously broken in the ground state, leading to only one gapless mode. The second mode now has a gap proportional to the magnetic field. The location of the distinct pole at ${\bf k}=X$ thus no longer scales to zero in the thermodynamic limit, but is simply equal the magnetic field. As mentioned in Sect.~\ref{sec:ms}, discussing the transverse moment $m_s$, quantum fluctuations initially decrease with increasing magnetic field and reach a minimum at around $m\approx 0.15$. It is therefore not surprising that the classical dispersion is almost identical to our exact results in the vicinity of these polarizations. Comparing the zone-boundary dispersions (from M to S) for different magnetizations shown in the first row of Fig.~\ref{fig:longall}, one finds that the zero-field effect is inverted and that for $m\gtrsim1/8$, the spin-wave energy at $(\pi,0)$ is higher than at $(\pi/2,\pi/2)$. At $m=1/8$, for instance, the energy differs by $\sim 5\%$, with a slope opposite to the one at zero field.

With increasing magnetization, the weights of the magnon poles decrease, while at the same time, poles at higher energies become more important. This modification is first observed at momenta close to $X=(\pi,\pi)$, but spreads quickly along almost the whole path shown in Fig.~\ref{fig:longall}. Only momenta around the gapless point $\Gamma=(0,0)$ have well defined spin-wave poles for all meaningful magnetizations. This is the finite-size manifestation of the magnon instability predicted in Ref.~\onlinecite{zhitomirsky99}. In the decaying regime, the lowest-lying poles have very small weight and are thus difficult to identify. As a guide to the eye, we have connected them in the case of the 64-sites cluster, see last row of Fig.~\ref{fig:longall}, showing the presence of multiple spin-wave continua below the primary spin-wave branch. The possibility to resolve poles with very little weight in the spectral functions is a conceptual advantage of exact diagonalization compared to other methods like quantum Monte-Carlo simulations. Using this information, it is possible to detect the availability of phase space for the decay of single spin waves, even in the case where the lifetimes are large.

Since our calculations were done on finite clusters, one must be careful interpreting results at fields very close to saturation, which might be obtained from small Hilbert spaces with only one or two flipped spins. These situations are clearly not representative for the infinite system and we thus only consider data from sectors with magnetizations $m \leq m_\text{max}=1/2-4/N$.

The magnon dispersion at $h=3.5$ is in very good agreement with recent quantum Monte-Carlo simulations combined with stochastic analytical continuation~\cite{syljuasen08}, compare second plot in the last row of Fig.~\ref{fig:longall} with Fig.~4 of Ref.~\onlinecite{syljuasen08}. Concerning the magnon decay, our exact diagonalizations indicate that spin waves close to $(\pi,\pi)$ have a very long, but finite lifetime. In contrast, the density plot presented in Ref.~\onlinecite{syljuasen08} suggests that spin waves are stable in this region. Around $(\pi,0)$, the situation is much clearer, and both methods predict a region of instability, despite the fact that the two-magnon decay kinematics of the spin-wave theory do not allow for an instability at this momentum. It would be interesting to investigate these subtle differences in more detail and compare the predications of these two numerical approaches to more advanced spin-wave calculations~\cite{zhitomirsky99}.

\subsection{Quantum fluctuations and finite-size effects}
\begin{figure*}
\centerline{\includegraphics[width=0.85\linewidth,clip]{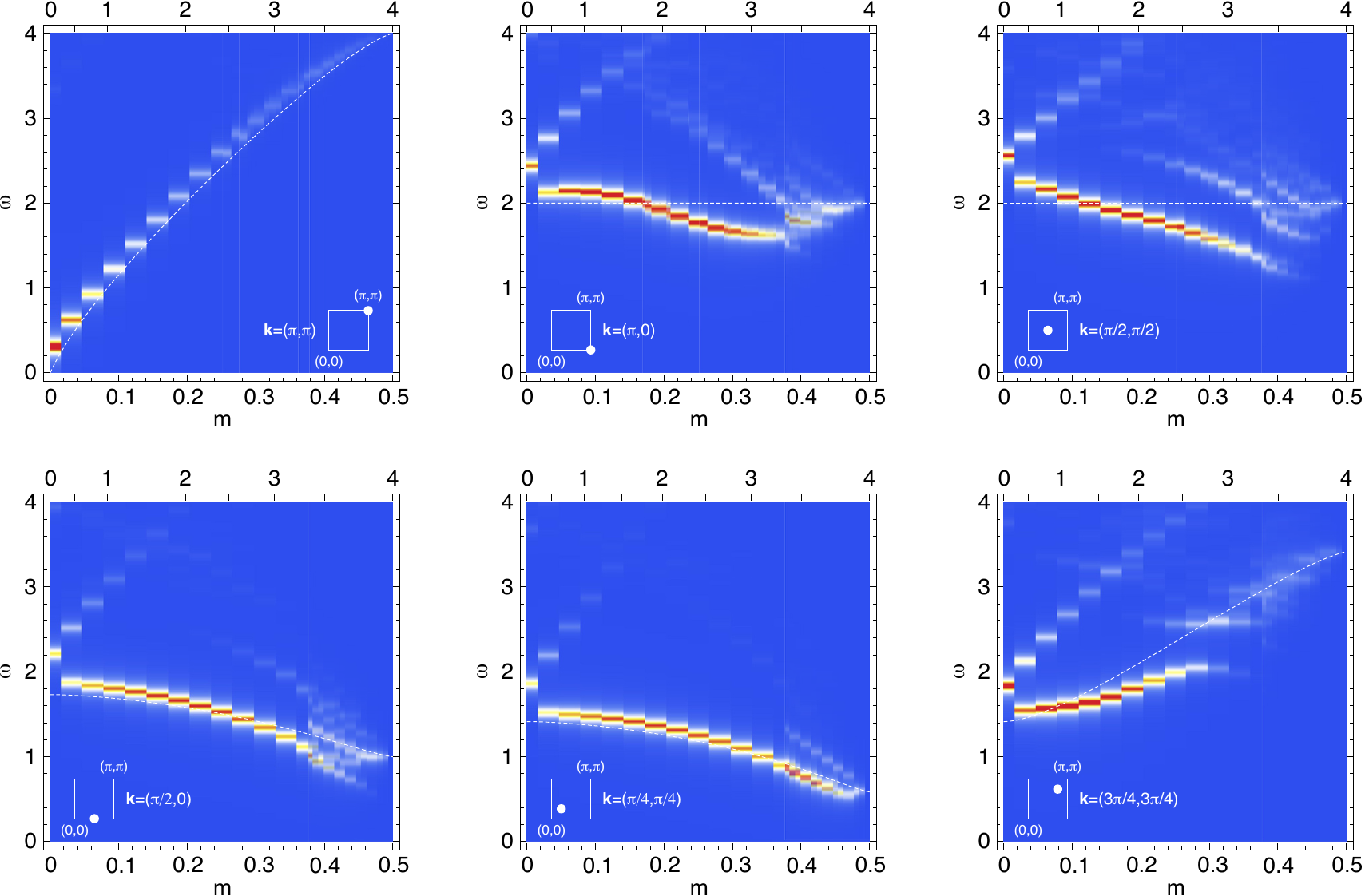}}
\caption{\emph{(Color online) Density plots of the longitudinal dynamical structure factors $S^{zz}({\bf k},\omega)$ for various momenta ${\bf  k}$. These plots are obtained from the raw data presented in Fig.~\ref{fig:dynsf}(a) after artificial broadening by an imaginary part $\eta=0.05$. Structure factors from all available clusters are superimposed. Dashed lines represent the dispersion obtained within linear spin-wave theory and the upper $x$-axes indicate the mapping of the magnetization onto the magnetic field.} 
\label{fig:dynszz}}
\end{figure*}

After this overview, let us now look at some of the discovered features in more detail, using the \emph{exact} dynamical structure factors \emph{without} any interpolation presented in Fig.~\ref{fig:dynsf}. A false-color density plot of the longitudinal structure factor is shown in Fig.~\ref{fig:dynszz} after artificial broadening by an imaginary part $\eta=0.05$. Because data from all available clusters is superimposed, some of the features abruptly appear as more Hilbert spaces become accessible for computations. In general, we expect the longitudinal spin correlations at momentum ${\bf k}$ to resemble the transverse ones at momentum ${\bf k}+{\bf Q}$, i.e., 
\begin{equation*}
S^{zz}(\omega,{\bf q})/S^{zz}({\bf q}) \approx S^{xy}(\omega,{\bf q}+{\bf Q})/S^{xy}({\bf q}+{\bf Q}) \ .
\end{equation*}
This follows from the tower of states shown in Fig.~\ref{fig:spectrum}. Given the similar structure of levels in neighboring sectors, it reasonable to expect the dynamical structure factors to reflect these similarities. The results displayed in Fig.~\ref{fig:dynsf} nicely support this hypothesis.

At zero field, it is well known that the dispersion obtained within lowest-order spin-wave theory is renormalized by quantum fluctuations, leading to energies roughly $16\%$ higher than this simplest prediction. As explained in Sect.~\ref{sec:ms}, discussing the ordered transverse moment $m_s$, quantum fluctuations are minimal at magnetizations around $m\approx0.15$. This is exactly the regime in which the lowest order spin-wave dispersion is in good agreement with exact numerical calculations. However, away from this point, there are sizable deviations. Similar to the zero-field situation, the most pronounced effect is observed at the magnetic zone boundary along $(\pi,0)$ and $(\pi/2,\pi/2)$, for which spin-wave calculations predict a flat energy $\omega=2 J$, independent of the magnetization. In contrast, our field dependent results at $(\pi/2,\pi/2)$ [and to a lesser extent also $(\pi,0)$] reveal a large negative slope, not at all compatible with the flat linear spin-wave results. Away from the magnetic zone boundary, the qualitative features of the field dependence are nevertheless nicely captured by the harmonic spin-wave spectrum. Going beyond linear spin-wave theory, one can include the first $1/S$ corrections in perturbation theory, see Eq.~9 in Ref.~\onlinecite{zhitomirsky99}. In Fig.~\ref{fig:dynsf}, these perturbative calculations are shown as solid lines. As explained in Ref.~\onlinecite{zhitomirsky99}, such a perturbative approach breaks down at larger fields, at which one should take into account the renormalized magnon energies and solve the Dyson equation self-consistently. For this reason, the solid lines do not extend over the whole field range. Given the rather good agreement between our exact numerical results and these perturbative energies, we conclude that the dominant dispersive features are well captured including $1/S$ corrections.

Our findings corroborate recent results revealing that noncollinear magnetic structures encounter sizable renormalization of the excitation spectrum 
when improving the linear (harmonic) spin-wave theory by the first $1/S$ corrections~\cite{chernyshev06}. While this happens already in zero field for the triangular lattice~\cite{zheng06,starykh06} with noncollinear 120$^\circ$ magnetic order, on the square lattice~\cite{zhitomirsky99}, this effect is driven by the magnetic field, and can thus be tuned in experiments.

In the longitudinal structure factors away from $(0,0)$ and $(\pi,\pi)$, one observes a sometimes sizable jump of the energy of the first pole as the system is polarized. This discontinuity is caused by different finite-size effects present at zero field and for $m>0$. The longitudinal magnon mode at $S^z=0$ is in fact the $S^z$$=$0-component of a spin $1$ level. At $m=0$, the energy thus contains a finite-size contribution of the Anderson tower that scales away as $1/N$. In contrast, at $m>0$, the longitudinal structure factor couples predominantly states of equal spin, and the leading $1/N$ finite-size corrections are absent.

Apart from the main spin-wave branch, a second distinct feature of the dynamical structure factors is the almost linearly increasing mode at low fields. Its intensity fades away as the field is increased and it is visible at all momenta. This is not a new mode, but a finite-size hybridization between the longitudinal and the transverse spin structure factors, which can be explained by the Wigner-Eckart theorem. 
At a magnetic field $h$, corresponding to a magnetization $m$, the ground state is in a total spin $S_m=m N$ state with polarization $S^z=S_m$. The $S^+$ scattering operator entering $S^{-+}$ couples only to total spin $S=S_m+1$ states in the spectrum because it is the $+1$ component of a spin $1$ tensor operator. The $S^z$ scattering operator entering $S^{zz}$ couples to total spin $S=S_m$ and $S=S_m+1$ states in the spectrum, while $S^-$ couples to total spin $S=S_m-1$, $S_m$ and $S_m+1$ states.
Using the Wigner-Eckart theorem one can prove that the part of the $S^z$-operator that couples to $|S_m+1,S_m\rangle$ states is actually related to the action of $S^+$, and the following relation holds
\begin{equation*}
S^{zz}(\mathbf q,\omega)|_{S=S_m+1} =  C^{-+}_{zz} \times S^{-+}(\mathbf{q},\omega+h),
\end{equation*}
with
\begin{equation*}
C^{-+}_{zz} =\left|\frac{\langle 1,0~;~S_m,S_m|S_m+1,S_m \rangle}{\langle 1,1~;~S_m,S_m|S_m+1,S_m+1 \rangle}\right|^2 = \frac{1}{S_m+1} \ ,
\end{equation*}
a squared fraction of Clebsch-Gordan coefficients. As one increases $S_m=Nm$, either by increasing $m$ at fixed system size $N$, or by increasing $N$ at fixed $m$, the proportionality factor $C^{-+}_{zz}$ decreases as $1/S_m$ for large $S_m$, and therefore, the shifted shadow of $S^{-+}$ fades away, unless $S^{-+}$ is sufficiently divergent. Similarly, one can show that $S^{+-}$ contains a shadow of $S^{zz}$ which is generically Clebsch-Gordan suppressed at large $m$ and $N$.

\begin{figure*}
\centerline{\includegraphics[width=0.95\linewidth,clip]{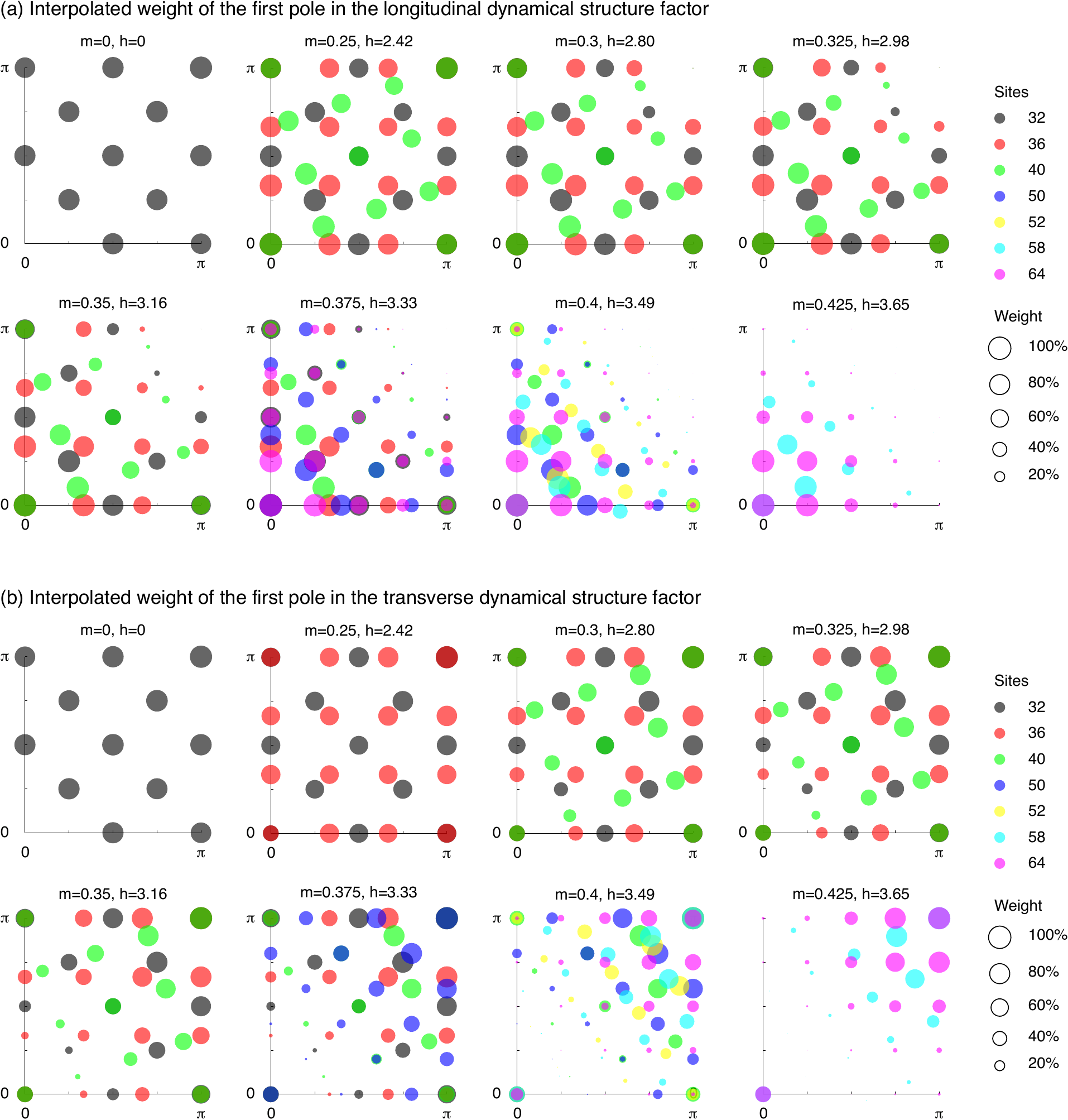}}
\caption{\emph{Interpolated weights of the first poles in the longitudinal (a) and transverse (b) dynamical structure factors for various magnetizations $m$. Poles with a sizable weight represent stable magnons with a long lifetime whereas small weights indicate that spin-wave energies lie within a continuum. Magnons thus have a short lifetime and are susceptible to spontaneous decay.} 
\label{fig:weights}}
\end{figure*}

\subsection{Magnon instabilities}
We have seen that the location of the dominant pole in the longitudinal structure factor at momentum $(\pi,\pi)$, and hence at $(0,0)$ in the transverse spin correlations, is equal to the magnetic field. The corresponding plots in Fig.~\ref{fig:dynsf} thus represent nothing but the inverse magnetization curve $m(h)$. In the longitudinal structure factor, the lowest-lying pole at small magnetizations is also the dominant one, but interestingly, for $m\geq9/32$, we observe an increasing number of poles with almost negligible weight appearing below the main spin-wave feature. Because of their low intensity, we have indicated the locations of these poles by uniformly sized open symbols in the area denoted ``zoom''. Only weights larger than $10^{-6}$ have been included in the plot, but most symbols represent poles with weights around $10^{-3}$. For small clusters, i.e., $N\leq36$, this group of levels first appears around $m\approx0.1$, still above the spin-wave band, remaining roughly constant at energy $\omega \approx 3$, until they cross the main magnon branch and lose some of their intensity. Data from bigger clusters suggests that these low-lying excitations form a continuum. Hence magnons in the main branch seem to have a very long but finite lifetime. Given the tiny weight of the low-energy excitations, one can conclude that while there is phase space available at these energies, the decay matrix element is extremely small. 
The dominant pole in the longitudinal $(\pi,\pi)$ structure factor in sector $S^z=S_m$ originates from the local copy of the ground state level in the neighboring spin sector with spin $S=S_m+1$. In contrast, the almost negligible poles can be traced back to low-lying spin $S=S_m$ states whose energies decreases with increasing magnetic field. The transverse structure factor at $(0,0)$ does not exhibit such a continuum of low-intensity poles below the spin-wave branch because the $S^-$ operator acts as a global spin lowering operator creating an exact excited eigenstate, which gives rise to a single delta function in $S^{xy}(\mathbf{q}=\Gamma,\omega)$.

Within linear spin-wave approximation, magnons are stable at any momentum and magnetic field, simply because the Hamiltonian does not contain terms that would allow a single magnon to decay into two spin waves. Knowing that such processes are however present in the original Hamiltonian, one can use purely kinematical arguments to study the necessary conditions for magnon instabilities. Following Ref.~\onlinecite{zhitomirsky99}, we expand the spin-wave dispersion in the vicinity of the gapless mode at $(\pi,\pi)$ leading to
\begin{equation*}
\epsilon_{\bf q} \approx c {q} \left[1+\alpha_{ q} \ {q}^2\right] \ ,
\end{equation*}
where $\left|{\bf q}-{\bf Q}\right| \ll 1$ and $c$ is the spin-wave velocity. The curvature $\alpha_{q}$ is wave-vector dependent. Along the diagonal, one finds $\alpha=-7/3+2/\sqrt{1-h^2}$, which changes sign at a critical field $h_c=2/\sqrt{7} h_s\ \approx 0.755 \ h_s$, corresponding to a magnetization $m_c \approx 0.33$. This signals the onset of the spin-wave instability~\cite{zhitomirsky99}. 

Away from the special $(\pi,\pi)$-point, our results are much clearer and show evidence for the decay of spin waves in multi spin-wave continua. Inspecting the longitudinal structure factor at $(3\pi/4,3\pi/4)$ for instance, or equivalently, the transverse spin correlations at $(\pi/4,\pi/4)$, we see that the spin-wave band fades away at $m\approx 0.3$ and enters a continuum, which on these finite clusters manifests itself as an area with densely distributed poles having relatively small intensity. For polarizations close to the ferromagnetic regime, the number of poles decreases rapidly. In the Hilbert space with only one flipped spin, the single pole coincides with the dashed classical line because in the limit $m\to1/2$, spin-wave results for the ferromagnet are exact. Compared to the situation at $(\pi,\pi)$, which seems to be unique, our results at $(3\pi/4,3\pi/4)$ clearly show that the initially well defined magnon branch dives into a continuum and hence acquires a finite lifetime. 

Looking at other points in the Brillouin zone, one clearly recognizes similar longitudinal instabilities at $(\pi/2,\pi/2)$, $(\pi,0)$, and $(\pi/2,0)$, or equivalently at $(\pi,\pi)$ shifted points in the transverse structure factor. The critical field at which magnon decay sets in depends on the momenta. To identify the regions of instability, it is useful to plot the weights of the first poles in the structure factors as a function of the magnetization. Since different clusters have incompatible magnetization steps, it is necessary to use an interpolation to collect data from all clusters at a given magnetization. Figs.~\ref{fig:weights} show the weights of the lowest-lying pole in the longitudinal and the transverse structure factors for various magnetizations $m$. To eliminate spurious effects arising from very small Hilbert spaces at high polarization, we have only included data from sectors with $m\leq m_\text{max}$. We would like to emphasize that this way of illustrating the instabilities with respect to spontaneous decay does not contain any information about the lifetime of the magnons, but merely indicates the availability of phase space for decay. It is also important to note that in contrast to Fig.~\ref{fig:dynsf}, representing the full structure factors, in Fig.~\ref{fig:weights}, we only indicate the weights of the poles with lowest energy. This plot nicely illustrates the spreading out of the decay region, starting around $(\pi,\pi)$ in the longitudinal and $(0,0)$ in the transverse structure factor, respectively. Note that this evolution is opposite to the instability evolution advocated in Ref.~\onlinecite{zhitomirsky99}, where the decay into two spin waves sets in in the vicinity of the gapless point. We believe that this behavior is difficult to study, given the infrared cutoff imposed by the small size limitations of exact diagonalizations. On the other hand, our results clearly demonstrate the importance of multi-spin-wave continua for the decay processes, and those are first visible in the opposite corner of the Brillouin zone.

Based on these results, we can say that high energy magnons around $(\pi,\pi)$ become unstable at a critical polarization $m_c \approx 0.3$. For higher magnetizations, the region of instability quickly increases. At $m=0.4$, we expect the spin waves in more than half the Brillouin zone to have a finite lifetime.

\section{Conclusion\label{sec:conclusion}}
We have studied the antiferromagnetic spin-$\frac{1}{2}$ Heisenberg model on the square lattice in a magnetic field. For the first time, we have calculated the field dependence of the microscopic parameters of the low-energy long-wavelength $\sigma$ model description, namely the spin-wave velocity $c$, the spin-stiffness $\rho_s$, and the transverse magnetic susceptibility $\chi_\perp$. The motivation behind this work being the availability of compounds well described by this prototypical model, the main part of this paper has been devoted to studying the field dependence of the dynamical spin structure factors, directly proportional to the neutron scattering cross section. On the one hand, we hope that our comprehensive presentation of these dynamical quantities encourages future theoretical and especially experimental work to examine particular aspects in even more detail. On the other hand, we are confident that our results obtained by means of extensive exact diagonalizations of finite clusters with up to 64 sites, for which no approximation or simplification has been made, serve as a benchmark for other methods.

\acknowledgments
We thank M.~E. Zhitomirsky and A.~L. Chernyshev for discussions and advice on the perturbative $1/S$ spin-wave corrections. We are grateful to A.~Honecker for providing us with the magnetization curves shown in Fig.~\ref{fig:m} and acknowledge the  allocation of computing time on the machines of the Swiss National Superconducting Centre (CSCS). This work was supported by the Swiss National Fund (SNF).


\begin{thebibliography}{10}

\bibitem{manousakis91}
E.~Manousakis,
\href{http://dx.doi.org/10.1103/RevModPhys.63.1}{Rev. Mod. Phys. \textbf{63}, 1
  (1991)}.

\bibitem{barnes91}
T.~Barnes,
\href{http://www.worldscinetarchives.com/cgi-bin/wsnetarchives/article.cgi?jn=%
ijmpc&vol=02&yr=1991&issue=0202&art=S0129183191000949.html}{Int. J. Mod. Phys.
  C \textbf{2}, 659 (1991)}.

\bibitem{vaknin87}
D.~Vaknin, S.~K. Sinha, D.~E. Moncton, D.~C. Johnston, J.~M. Newsam, C.~R.
  Safinya, and H.~E. King, Jr.,
\href{http://link.aps.org/doi/10.1103/PhysRevLett.58.2802}{Phys. Rev. Lett.
  \textbf{58}, 2802 (1987)}.

\bibitem{coldea01}
R.~Coldea, S.~M. Hayden, G.~Aeppli, T.~G. Perring, C.~D. Frost, T.~E. Mason,
  S.-W. Cheong, and Z.~Fisk,
\href{http://link.aps.org/doi/10.1103/PhysRevLett.86.5377}{Phys. Rev. Lett.
  \textbf{86}, 5377 (2001)}.

\bibitem{birgeneau99}
R.~J. Birgeneau, M.~Greven, M.~A. Kastner, Y.~S. Lee, B.~O. Wells, Y.~Endoh,
  K.~Yamada, and G.~Shirane,
\href{http://link.aps.org/doi/10.1103/PhysRevB.59.13788}{Phys. Rev. B
  \textbf{59}, 13788 (1999)}.

\bibitem{elstner95}
N.~Elstner, A.~Sokol, R.~R.~P. Singh, M.~Greven, and R.~J. Birgeneau,
\href{http://link.aps.org/doi/10.1103/PhysRevLett.75.938}{Phys. Rev. Lett.
  \textbf{75}, 938 (1995)}.

\bibitem{cuccoli97}
A.~Cuccoli, V.~Tognetti, R.~Vaia, and P.~Verrucchi,
\href{http://link.aps.org/doi/10.1103/PhysRevB.56.14456}{Phys. Rev. B
  \textbf{56}, 14456 (1997)}.

\bibitem{anderson52}
P.~W. Anderson,
\href{http://link.aps.org/doi/10.1103/PhysRev.86.694}{Phys. Rev. \textbf{86},
  694 (1952)}.

\bibitem{oguchi60}
T.~Oguchi,
\href{http://link.aps.org/doi/10.1103/PhysRev.117.117}{Phys. Rev. \textbf{117},
  117 (1960)}.

\bibitem{chakravarty89}
S.~Chakravarty, B.~I. Halperin, and D.~R. Nelson,
\href{http://dx.doi.org/10.1103/PhysRevB.39.2344}{Phys. Rev. B \textbf{39},
  2344 (1989)}.

\bibitem{kim99}
Y.~J. Kim, A.~Aharony, R.~J. Birgeneau, F.~C. Chou, O.~Entin-Wohlman, R.~W.
  Erwin, M.~Greven, A.~B. Harris, M.~A. Kastner, I.~Ya. Korenblit, Y.~S. Lee,
  and G.~Shirane,
\href{http://link.aps.org/doi/10.1103/PhysRevLett.83.852}{Phys. Rev. Lett.
  \textbf{83}, 852 (1999)}.

\bibitem{sandvik01}
A.~W. Sandvik and R.~R.~P. Singh,
\href{http://link.aps.org/doi/10.1103/PhysRevLett.86.528}{Phys. Rev. Lett.
  \textbf{86}, 528 (2001)}.

\bibitem{ronnow01}
H.~M. R{\o}nnow, D.~F. McMorrow, R.~Coldea, A.~Harrison, I.~D. Youngson, T.~G.
  Perring, G.~Aeppli, O.~F. Sylju{\aa}sen, K.~Lefmann, and C.~Rischel,
\href{http://link.aps.org/doi/10.1103/PhysRevLett.87.037202}{Phys. Rev. Lett.
  \textbf{87}, 037202 (2001)}.

\bibitem{christensen04}
N.~B. Christensen, D.~F. McMorrow, H.~M. R{\o}nnow, A.~Harrison, T.~G. Perring,
  and R.~Coldea,
\href{http://linkinghub.elsevier.com/retrieve/pii/S030488530400068X}{J. Magn.
  Magn. Mater. \textbf{272-276}, 896 (2004)}.

\bibitem{christensen07}
N.~B. Christensen, H.~M. R{\o}nnow, D.~F. McMorrow, A.~Harrison, T.~G. Perring,
  M.~Enderle, R.~Coldea, L.~P. Regnault, and G.~Aeppli,
\href{http://www.pnas.org/cgi/content/abstract/104/39/15264}{Proc. Natl. Acad.
  Sci. USA \textbf{104}, 15264 (2007)}.

\bibitem{woodward02}
F.~M. Woodward, A.~S. Albrecht, C.~M. Wynn, C.~P. Landee, and M.~M. Turnbull,
\href{http://link.aps.org/doi/10.1103/PhysRevB.65.144412}{Phys. Rev. B
  \textbf{65}, 144412 (2002)}.

\bibitem{lancaster07}
T.~Lancaster, S.~J. Blundell, M.~L. Brooks, P.~J. Baker, F.~L. Pratt, J.~L.
  Manson, M.~M. Conner, F.~Xiao, C.~P. Landee, F.~A. Chaves, S.~Soriano, M.~A.
  Novak, T.~P. Papageorgiou, A.~D. Bianchi, T.~Hermannsd{\"o}rfer, J.~Wosnitza,
  and J.~A. Schlueter,
\href{http://link.aps.org/doi/10.1103/PhysRevB.75.094421}{Phys. Rev. B
  \textbf{75}, 094421 (2007)}.

\bibitem{coomer07}
F.~C. Coomer, V.~Bondah-Jagalu, K.~J. Grant, A.~Harrison, G.~J. McIntyre, H.~M.
  R{\o}nnow, R.~Feyerherm, T.~Wand, M.~Mei{\ss}ner, D.~Visser, and D.~F.
  McMorrow,
\href{http://link.aps.org/doi/10.1103/PhysRevB.75.094424}{Phys. Rev. B
  \textbf{75}, 094424 (2007)}.

\bibitem{osano82}
K.~Osano, H.~Shiba, and Y.~Endoh,
\href{http://ptp.ipap.jp/link?PTP/67/995}{Prog. Theor. Phys. \textbf{67}, 995
  (1982)}.

\bibitem{zhitomirsky99}
M.~E. Zhitomirsky and A.~L. Chernyshev,
\href{http://link.aps.org/doi/10.1103/PhysRevLett.82.4536}{Phys. Rev. Lett.
  \textbf{82}, 4536 (1999)}.

\bibitem{fisher89}
D.~S. Fisher,
\href{http://dx.doi.org/10.1103/PhysRevB.39.11783}{Phys. Rev. B \textbf{39},
  11783 (1989)}.

\bibitem{halperin69}
B.~I. Halperin and P.~C. Hohenberg,
\href{http://link.aps.org/doi/10.1103/PhysRev.188.898}{Phys. Rev. \textbf{188},
  898 (1969)}.

\bibitem{igarashi92}
J.~Igarashi,
\href{http://link.aps.org/doi/10.1103/PhysRevB.46.10763}{Phys. Rev. B
  \textbf{46}, 10763 (1992)}.

\bibitem{sandvik97}
A.~W. Sandvik,
\href{http://link.aps.org/doi/10.1103/PhysRevB.56.11678}{Phys. Rev. B
  \textbf{56}, 11678 (1997)}.

\bibitem{neuberger89}
H.~Neuberger and T.~Ziman,
\href{http://link.aps.org/doi/10.1103/PhysRevB.39.2608}{Phys. Rev. B
  \textbf{39}, 2608 (1989)}.

\bibitem{hasenfratz90}
P.~Hasenfratz and H.~Leutwyler,
\href{http://dx.doi.org/10.1016/0550-3213(90)90603-B}{Nucl. Phys. B
  \textbf{343}, 241 (1990)}.

\bibitem{hasenfratz93}
P.~Hasenfratz and F.~Niedermayer,
\href{http://www.springerlink.com/content/v152426638l2m823}{Z. Phys. B
  \textbf{92}, 91 (1993)}.

\bibitem{sandvik99}
A.~W. Sandvik,
\href{http://link.aps.org/doi/10.1103/PhysRevB.59.R14157}{Phys. Rev. B
  \textbf{59}, 14157(R) (1999)}.

\bibitem{alet05}
F.~Alet, S.~Wessel, and M.~Troyer,
\href{http://link.aps.org/doi/10.1103/PhysRevE.71.036706}{Phys. Rev. E
  \textbf{71}, 036706 (2005)}.

\bibitem{alps07}
A.~F. Albuquerque, F.~Alet, P.~Corboz, P.~Dayal, A.~Feiguin, S.~Fuchs,
  L.~Gamper, E.~Gull, S.~G{\"u}rtler, A.~Honecker, R.~Igarashi, M.~K{\"o}rner,
  A.~Kozhevnikov, A.~M. L{\"a}uchli, S.~R. Manmana, M.~Matsumoto, I.~P.
  McCulloch, F.~Michel, R.~M. Noack, G.~Pawlowski, L.~Pollet, T.~Pruschke,
  U.~Schollw{\"o}ck, S.~Todo, S.~Trebst, M.~Troyer, P.~Werner, and S.~Wessel,
\href{http://linkinghub.elsevier.com/retrieve/pii/S0304885306014983}{J. Magn.
  Magn. Mater. \textbf{310}, 1187 (2007)}.

\bibitem{richter04}
J.~Richter, J.~Schulenburg, and A.~Honecker,
\href{http://www.springerlink.com/index/CCV22WCN1JNRXFHU.pdf}{Lecture Notes in
  Physics \textbf{645}, 85 (2004)}.

\bibitem{sandvik99b}
A.~W. Sandvik and C.~J. Hamer,
\href{http://link.aps.org/doi/10.1103/PhysRevB.60.6588}{Phys. Rev. B
  \textbf{60}, 6588 (1999)}.

\bibitem{zhitomirsky98}
M.~E. Zhitomirsky and T.~Nikuni,
\href{http://link.aps.org/doi/10.1103/PhysRevB.57.5013}{Phys. Rev. B
  \textbf{57}, 5013 (1998)}.

\bibitem{gluzman93}
S.~Gluzman,
\href{http://www.springerlink.com/content/g9n6l78603946k55/?p=1cfda7e483294a07%
a04feb9f0cbb3560&pi=9%0A}{Z. Phys. B \textbf{90}, 313 (1993)}.

\bibitem{sachdev94}
S.~Sachdev, T.~Senthil, and R.~Shankar,
\href{http://link.aps.org/doi/10.1103/PhysRevB.50.258}{Phys. Rev. B
  \textbf{50}, 258 (1994)}.

\bibitem{honecker99}
A.~Honecker,
\href{http://www.iop.org/EJ/abstract/0953-8984/11/24/311}{J. Phys.: Condens.
  Matter \textbf{11}, 4697 (1999)}.

\bibitem{dzhaparidze78}
G.~I. Dzhaparidze and A.~A. Nersesyan,
\href{http://www.jetpletters.ac.ru/ps/1549/article_23715.shtml}{Pis'ma Zh. Eksp. Teor. Fiz. \textbf{27}, 356 (1978) 
[JETP Lett. \textbf{27}, 334 (1978)]}.

\bibitem{chernyshev09}
A.~L. Chernyshev and M.~E. Zhitomirsky,
\href{http://arxiv.org/abs/0902.4455v1}{arXiv:0902.4455v1 (unpublished)}.

\bibitem{einarsson95}
T.~Einarsson and H.~J. Schulz,
\href{http://link.aps.org/doi/10.1103/PhysRevB.51.6151}{Phys. Rev. B
  \textbf{51}, 6151 (1995)}.

\bibitem{pollock84}
E.~Pollock and D.~M. Ceperley,
\href{http://link.aps.org/doi/10.1103/PhysRevB.30.2555}{Phys. Rev. B
  \textbf{30}, 2555 (1984)}.

\bibitem{schulz96}
H.~J. Schulz, T.~Ziman, and D.~Poilblanc,
\href{http://jp1.journaldephysique.org/index.php?option=article&access=standar%
d&Itemid=129&url=/articles/jp1/abs/1996/05/jp1v6p675/jp1v6p675.html}{J. Phys. I
  \textbf{6}, 675 (1996)}.

\bibitem{kreisel08}
A.~Kreisel, F.~Sauli, N.~Hasselmann, and P.~Kopietz,
\href{http://link.aps.org/doi/10.1103/PhysRevB.78.035127}{Phys. Rev. B
  \textbf{78}, 035127 (2008)}.

\bibitem{hamer94}
C.~J. Hamer, W.~Zheng, and J.~Oitmaa,
\href{http://link.aps.org/doi/10.1103/PhysRevB.50.6877}{Phys. Rev. B
  \textbf{50}, 6877 (1994)}.

\bibitem{yang97}
M.~S. Yang and K.~H. M{\"u}tter,
\href{http://www.springerlink.com/index/Y1JVPN6GJX6KA2Y1.pdf}{Z. Phys. B
  \textbf{104}, 117 (1997)}.

\bibitem{syljuasen08}
O.~F. Sylju{\aa}sen,
\href{http://link.aps.org/doi/10.1103/PhysRevB.78.180413}{Phys. Rev. B
  \textbf{78}, 180413 (2008)}.

\bibitem{spremo06}
I.~Spremo,
\href{http://deposit.ddb.de/cgi-bin/dokserv?idn=981056474&dok_var=d1&dok_ext=p%
df&filename=981056474.pdf}{PhD Thesis  (2006)}.

\bibitem{bernu92}
B.~Bernu, C.~Lhuillier, and L.~Pierre,
\href{http://link.aps.org/doi/10.1103/PhysRevLett.69.2590}{Phys. Rev. Lett.
  \textbf{69}, 2590 (1992)}.

\bibitem{gagliano87}
E.~R. Gagliano and C.~A. Balseiro,
\href{http://link.aps.org/doi/10.1103/PhysRevLett.59.2999}{Phys. Rev. Lett.
  \textbf{59}, 2999 (1987)}.

\bibitem{dagotto94}
E.~Dagotto,
\href{http://dx.doi.org/10.1103/RevModPhys.66.763}{Rev. Mod. Phys. \textbf{66},
  763 (1994)}.

\bibitem{zheng05}
W.~Zheng, J.~Oitmaa, and C.~J. Hamer,
\href{http://link.aps.org/doi/10.1103/PhysRevB.71.184440}{Phys. Rev. B
  \textbf{71}, 184440 (2005)}.

\bibitem{chernyshev06}
A.~L. Chernyshev and M.~E. Zhitomirsky,
\href{http://link.aps.org/doi/10.1103/PhysRevLett.97.207202}{Phys. Rev. Lett.
  \textbf{97}, 207202 (2006)}.

\bibitem{zheng06}
W.~Zheng, J.~O. Fj{\ae}restad, R.~R.~P. Singh, R.~H. McKenzie, and R.~Coldea,
\href{http://link.aps.org/doi/10.1103/PhysRevLett.96.057201}{Phys. Rev. Lett.
  \textbf{96}, 057201 (2006)}.

\bibitem{starykh06}
O.~A. Starykh, A.~V. Chubukov, and A.~G. Abanov,
\href{http://link.aps.org/doi/10.1103/PhysRevB.74.180403}{Phys. Rev. B
  \textbf{74}, 180403 (2006)}.

\end{thebibliography}
\end{document}